\documentclass[
superscriptaddress,
reprint,
 amsmath,amssymb,
 aps, 
pre]{revtex4-2}

\usepackage[british]{babel}
\usepackage[utf8]{inputenc}
\usepackage[T1]{fontenc}

\expandafter\let\csname equation*\endcsname\relax
\expandafter\let\csname endequation*\endcsname\relax
\usepackage{amssymb,amsfonts}
\usepackage{physics}
\usepackage{graphicx}
\usepackage{float}
\usepackage{xurl}
\usepackage{wrapfig}
\usepackage{gensymb}
\usepackage{caption}
\usepackage{subcaption}
\usepackage{footnote}
\makesavenoteenv{tabular}
\usepackage{wasysym}
\usepackage{mathtools}
\usepackage{bm} %put text in bold in mathmode
\usepackage{dsfont}
\usepackage{soul}
\usepackage[dvipsnames]{xcolor}
\usepackage[pdftex, pdftitle={Article}, pdfauthor={Author}]{hyperref} % For hyperlinks in the PDF

\newcommand{\lams}{\ensuremath{\lambda_{\rm s}}} % stability exponent
\newcommand{\laml}{\ensuremath{\lambda_{\rm L}}} % Lyapunov exponent
\newcommand{\qcross}{ \ensuremath{q_{ \rm c }}}  % crossover position qc

\usepackage{subfiles} % Best loaded last in the preamble
\begin{document}
\title{Scrambling at the genesis of chaos}

\author{Thomas R. Michel}
\email{Contact author: t.michel@uliege.be}
\affiliation{CESAM Research Unit, University of Liège, 4000 Liège, Belgium}

\author{Mathias Steinhuber}
\affiliation{Institut für Theoretische Physik, Universität Regensburg, Regensburg, 93040 Germany}

\author{Juan Diego Urbina}
\affiliation{Institut für Theoretische Physik, Universität Regensburg, Regensburg, 93040 Germany}

\author{Peter Schlagheck}
\affiliation{CESAM Research Unit, University of Liège, 4000 Liège, Belgium}

\date{} % Leave empty to omit a date

\begin{abstract}
The presence of chaos in classical Hamiltonian  systems is witnessed by its maximal Lyapunov exponent, that quantifies the instability of motion through the exponential growth of indicators such as the trace of the stability matrix or the out-of-time-ordered correlator. On the other hand, integrable dynamics near unstable fixed points, which are in turn characterized by a stability exponent, can also induce such exponential growth. Following the paradigm of integrability-breaking as driven by nonlinear resonances that hallmarks the genesis of chaos, the integrability-chaos transition is universally described by a periodic perturbation applied to a generic pendulum. Remarkably, this means that within the corresponding separatrix dynamics, which is an unavoidable a consequence of the resonance scenario, both instability exponents must play a role as both dynamical regimes coexist. We report here the universality of the transition from instability to Lyapunov exponents, thus completing the resonance scenario at the level of indicators based on exponential growth. To achieve this goal we obtain an analytical expression for the time evolution near separatrices, which enables us to derive an analytical expression for the exponent that characterises chaos and its transition from local instability to global chaos. We support our claim for  the universality of this mechanism by studying two paradigmatic examples of the integrability-to-chaos transition, namely the kicked rotor and the driven pendulum.
\end{abstract}

\maketitle

\section{Introduction} \label{sec_intro}

% =============================================================================================
Chaos has been an active domain of research for many decades \cite{Ott,Tabor_1989,Lichtenberg_Lieberman_2013}. 
While being initially a classical concept, it was extended to quantum systems \cite{Gutzwiller,Stockmann_1999,Haake_2001}.
The notion of initially close trajectories that deviate exponentially from one another due to nonlinear dynamics is not directly applicable here since quantum systems are governed by a perfectly linear equation (the Schr\"odinger equation in the context of nonrelativistic quantum matter).
Indirect quantum signatures of classical chaos were thus investigated in the first place, among them, most prominently, the statistics of the energy-level spacings \cite{BGS} as well as the phase-space distribution of eigenstates \cite{Deutsch_1991,Srednicki_1994}.

More direct indicators of the Lyapunov instability that is characteristic of classical chaos can nevertheless be established for quantum systems, namely in terms of the Loschmidt echo \cite{Peres84,Jalabert_Pastawski_2001,Yan_Cincio_Zurek_2020,PG_Madhok_Lakshminarayan_2021,Bhattacharyya_Chemissany_Haque_Yan_2022} and of out-of-time-ordered correlators (OTOC) \cite{LarOvc69, Swingle_2018,Garcia_Bu_Jaffe_2023,Maldacena_Shenker_Stanford_2016,Rammensee_Urbina_Richter_2018, Yan_Cincio_Zurek_2020,PG_Madhok_Lakshminarayan_2021,Bhattacharyya_Chemissany_Haque_Yan_2022, Xu_Swingle_2024}.
The Loschmidt echo tracks the overlap of two wavefunctions that evolve with slightly different Hamiltonians starting from the same initial state. Chaotic dynamics translate into an exponential decrease of the echo, highlighting the unstable nature of the system. An OTOC is typically given by the square modulus of the commutator between a time evolving system observable and another one that remains stationary, whose expectation value with respect to a fixed initial state is monitored as a function of the evolution time. This quantum object is shown to grow exponentially in the presence of classical chaos, the exponential growth rate being given by twice the Lyapunov exponent of the classical dynamics \cite{Maldacena_Shenker_Stanford_2016, Rozenbaum_Ganeshan_Galitski_2017, Rammensee_Urbina_Richter_2018,Jalabert18,Michel_Urbina_Schlagheck_2025}.
As for the Loschmidt echo \cite{Peres84,Jalabert_Pastawski_2001}, saturation of the indicator occurs at some point (beyond the Ehrenfest time), due to the quantum nature of the system, being specifically characterized by a finite extent of the Planck cell \cite{Rammensee_Urbina_Richter_2018,Michel_Urbina_Schlagheck_2025}.

However, an exponential growth of the OTOC, when being observable (i.e., if the system under consideration is not too far away from the semiclassical regime characterized by small Planck cells), is not a sufficient indicator to ascertain the presence of classical chaos \cite{Pappalardi18,HummelGeiger,Rozenbaum_Bunimovich_Galitski_2020}, especially not if the state under consideration (with respect to which the OTOC expectation value is taken) is tightly localized about a specific point in phase space.
If that particular point happens to be an unstable (more precisely, hyperbolic) fixed point of the classical dynamics within an integrable system, the resulting scrambling dynamics along the associated separatrix structure also results in an exponential growth of the OTOC, the growth rate being this time related to the stability exponent $\lambda_\text{s}$ of the fixed point \cite{HummelGeiger,Hashimoto_Huh_Kim_Watanabe_2020,Morita_2022, Meier_Steinhuber_Urbina_Waltner_Guhr_2023}.
Intricate growth regimes, with growth rates crossing over from $2\lambda_{\text{s}}$ to $\lambda_{\text{s}}$, followed by an oscillatory behaviour at long times, were reported in
\cite{Steinhuber_Schlagheck_Urbina_Richter_2023} for integrable systems, thus indicating that the latter nevertheless markedly distinguish themselves from their chaotic counterparts. Quite obviously, unstable fixed points also occur in nonintegrable systems of all possible sorts, ranging from the near-integrable to the fully chaotic regime.
This raises a question of fundamental nature: if an OTOC observable is evaluated with respect to an initial state that is strongly localized about such an unstable fixed point in a chaotic system, how will the two reported OTOC growth mechanisms compete with each other? Will one of them dominate the other one (and if so, which one in which regime) or will there be some sort of ``merging'' of these two OTOC growth laws?

To address this question, we mainly focus here on the periodically modulated pendulum in the near-integrable limit, which can be considered as a standard model for integrability breaking in the vicinity of generic nonlinear resonances \cite{Lichtenberg_Lieberman_2013} and which allows for analytical investigations using similar approaches as for the Melnikov-Arnold integral \cite{Melnikov_1962,Arnold_1964,Chirikov_1979}.
The scrambling observable that we consider in this context is the mean trace of the stability matrix \cite{Tabor_1989} which can be considered to be of more fundamental nature than any OTOC observable (the latter always depending on the choice of individual system operators), even though it is purely classical and does, to our knowledge, not have a direct quantum counterpart.
We find that at the genesis of chaos the trace of the stability matrix exhibits an exponential growth with a rate, which we call the \textit{effective exponent}, that lies between the stability exponent of the unstable fixed point of the pendulum and the Lyapunov exponent characterizing the tiny chaotic separatrix layer in the near-integrable regime \cite{Meier_Steinhuber_Urbina_Waltner_Guhr_2023}. 
Our analytical findings are confirmed by numerical simulations, which furthermore show the equivalence (up to a factor 2) of OTOC and stability matrix effective exponents, which explore what happens in mixed regular-chaotic parameter regimes, and whichdemonstrate the universality of our findings for generic driven or kicked one-degree-of-freedom systems.

This paper is structured as follows. In Section \ref{sec_instab} we introduce in more detail the concept of OTOCs and their classical counterparts, and provide a first numerical survey on OTOC growth behaviors associated with unstable fixed points in nonintegrable systems. 
In Section \ref{sec_analyticalDescription}, we establish an approximate analytical description of the driven pendulum, which is formally valid in the limit of weak perturbations from integrability. 
We then show in Section \ref{sec_expGrowthFromTr} how to analytically extract an approximate expression for the effective exponent characterizing the growth of the trace of the stability matrix. 
Section \ref{sec_compl} is devoted to complementary numerical investigations, where we show how our findings are universal and can be observed in other systems as well, in particular also using OTOCs instead of stabilitiy matrix traces. 
We also numerically investigate in that section %\ref{sec_mixedAndHierarchies} 
what happens farther away from integrability as well as for very long evolution times.
\section{Instability indicators}
\label{sec_instab}
Let us introduce in more details the probes of instability that we use. The OTOC is defined as the expectation value of the square modulus of a commutator between two local operators $\hat{A}$ and $\hat{B}$:
\begin{align} \label{OTOC_def}
    C(t) = \Tr\left( \rho \abs{\left[\hat{A}(t), \hat{B}(0) \right]}^2 \right).
\end{align}
where $\rho$ is the density matrix of the system, with here $\rho =\ket{\psi}\bra{\psi}$ as we consider pure states, with $\ket{\psi}$ the initial state of the system, and $\hat{A}(t)$ is the time-evolved operator $\hat{A}$ using the Heisenberg representation, i.e.
\begin{align} \label{A_t_operator}
    \hat{A}(t) =  \hat{U}^\dagger(t) \hat{A} \hat{U}(t),
\end{align}
with $\hat{U}(t)$ the evolution operator \cite{Rammensee_Urbina_Richter_2018}. We assume the operators to be hermitian in order to simplify the following.\\
\indent The OTOC can be thought as probing the spreading of operator $\hat{A}$ using $\hat{B}$ \cite{Swingle_2018}. Initially, it was thought that if the OTOC presents an exponential growth, it means the system was chaotic in a quantum sense. This was further supported by the fact that if the system admitted a classical limit, the classical dynamics would be chaotic \cite{Jalabert_Pastawski_2001}, and the exponential growth rate of the classical OTOC, equal for short time to the quantum one, would be the classical Lyapunov exponent \cite{Rozenbaum_Ganeshan_Galitski_2017, Rammensee_Urbina_Richter_2018}. A bridge was then made between quantum and classical chaos. \\
\indent Using either Wigner-Moyal \cite{Rammensee_Urbina_Richter_2018, Richter_Urbina_Tomsovic_2022} or the van Vleck-Gutziller propagator and the diagonal approximation \cite{Michel_Urbina_Schlagheck_2025}, it can be shown that in the classical limit $\hbar \to 0$ the OTOC becomes a Poisson bracket:
\begin{align}
    \bra{\psi} \abs{\left[\hat{A}(t), \hat{B} \right]}^2 \ket{\psi} = &\hbar^2 \int\dd \bm{q} \dd\bm{p}\ \abs{\{ A(\bm{q},\bm{p},t), B(\bm{q},\bm{p})\}}^2 \nonumber\\
    & \times W(\bm{q},\bm{p}) +\mathcal{O}(\hbar^3)
    \end{align}
with $A$, $B$ the Weyl symbol of operators $\hat{A}$ and $\hat{B}$, $W$ the Wigner function of the initial state $\ket{\psi}$,  $\bm{q}=(q_1,...,q_L)$ and $\bm{p}=(p_1,...,p_L)$ the position and momentum in phase space \cite{Polkovnikov_2010}. It can be made apparent that the OTOC follows an exponential growth in the presence of an instability. Considering $\hat{A}=\hat{q}_i$, $\hat{B} = \hat{p}_j$, the OTOC becomes
\begin{align}
       \bra{\psi} \abs{\left[\hat{q}_i(t), \hat{p}_j \right]}^2 \ket{\psi} & = \hbar^2 \int\dd \bm{q} \dd\bm{p}\ \left( \frac{\partial q_i}{\partial q_j}(t)  \right)^2 W(\bm{q},\bm{p})\nonumber \\
      & \ \ \ +\mathcal{O}(\hbar^3) \ \propto \hbar^2 e^{2\lambda t} \label{eq_otoc_classical}
\end{align}
with $\frac{\partial q_i}{\partial q_j}(t)$ an element of the stability matrix, which follows an exponential law in the presence of an instability, and $\lambda$ corresponding to the (classical) Lyapunov exponent of the system if the system is chaotic. This limit is valid only up to the Ehrenfest time, after which the quantum OTOC saturates due to scrambling \cite{Rammensee_Urbina_Richter_2018}.

\begin{figure}
	\centerline{
		\includegraphics[width=0.5\textwidth, angle=0, trim={0cm 0cm 0cm 0cm}, clip]{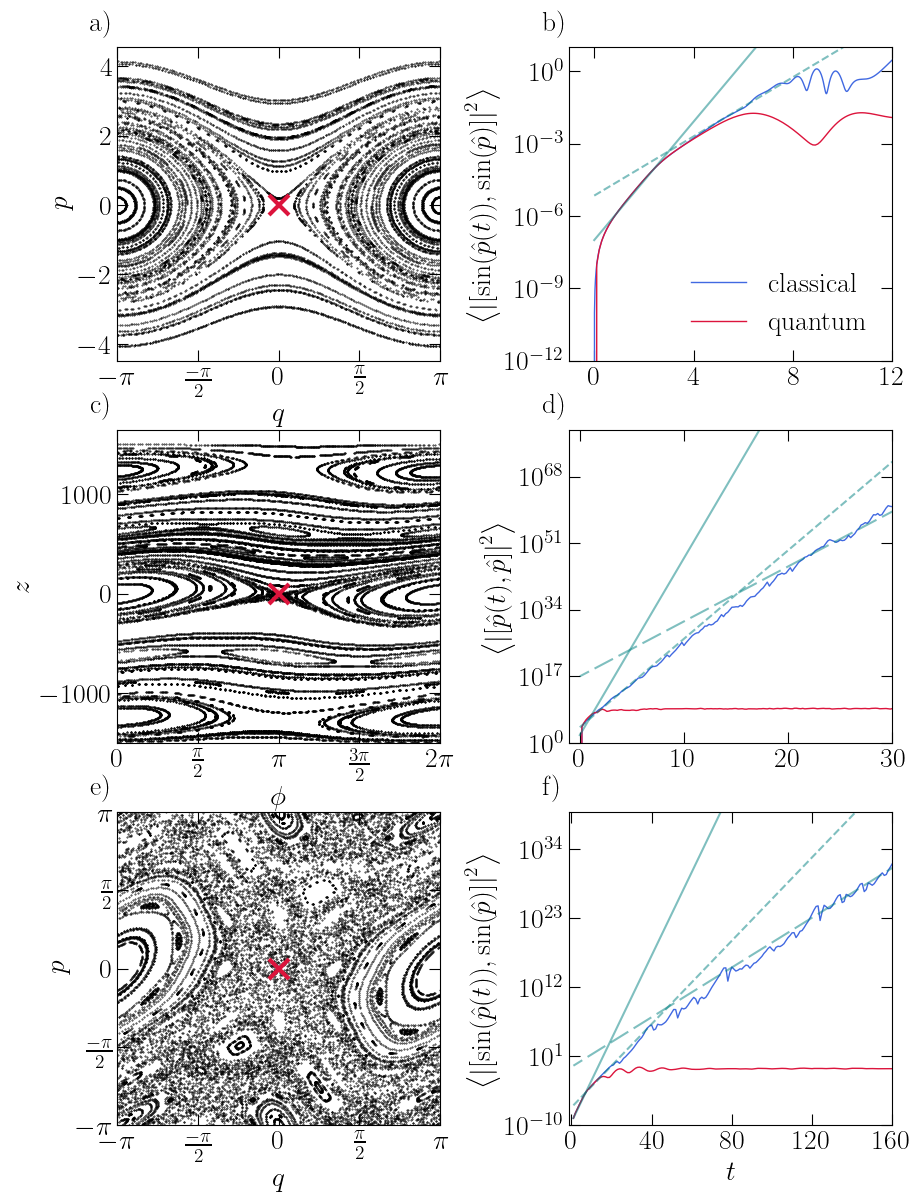}
	}
	\caption{OTOC growth in integrable, near-integrable and mixed systems. \textit{Left column}: Phase-space representations of a) the simple pendulum given by \eqref{eq_hamiltonian} with $\kappa=0$, c) the kicked Bose-Hubbard dimer given by \eqref{eq_kickedDimer} 100 particles, $U=J$, and the period and strength of the hopping kick $\tau = 0.2 J$ and e) the kicked rotor given by\eqref{eq_kickedRotor} with kicking parameter $K=0.6$. The center of the initial coherent state is indicated by a cross and corresponds to an unstable fixed point.
    \textit{Right column}: quantum (red) and classical (blue) OTOC for b) the simple pendulum with $\hbar=2^{-11}$, d) the kicked dimer with $\hbar=1/3000$ and f) the kicked rotor with $\hbar=2^{-14}$. The operators used for the OTOC are shown in the label of the vertical axis. The straight lines represent exponential growth, with the rate given by $2\lams$ (cyan), $\lams$ (dashed cyan), and $(\lams+\lambda_{\text{L}})/2$ (long-dashed cyan).}
    \label{fig_vitrine}
\end{figure}
However, it is recently known that the OTOC can exhibit an exponential regime even in the absence of chaos. This occurs if the initial state is located at a local instability, for instance a hyperbolic fixed point, where the exponent in \eqref{eq_otoc_classical} corresponds then to the stability exponent $\lams$. In that case, before the saturation the OTOC will undergo a crossover to a $\propto e^{\lams t}$ regime \cite{Steinhuber_Schlagheck_Urbina_Richter_2023}.

In this work, we will consider regimes between the integrable case and the completely chaotic one, which are all represented in Figure \ref{fig_vitrine}. 
We find that an effective exponent, $\lambda_{\text{eff}}$, whose value is between the stability exponent and the Lyapunov exponent plays a role in the transition from the regular to the chaotic case. 
To highlight the universality of our findings, each row corresponds to a different system. The first column shows the phase-space representation of a system with a hyperbolic fixed point at the centre, while the second one shows the quantum and classical OTOC for an initial states centred on that hyperbolic fixed point. The first row shows the integrable case for a simple pendulum. The $2\lams \to \lams$ transition is highlighted by representing exponentials using straight lines, the solid blue being proportional to $e^{2 \lams t}$ and the dashed one to $e^{\lams t}$. The quantum OTOC then saturates whereas the classical one continues to grow. The second row shows a quasi-integrable case, featuring dominantly regular dynamics with a thin chaotic layer around the hyperbolic fixed point and the homoclinic manifold linked to it. Here, there is an additional and new crossover where the exponential growth rate goes from $\lams$ to twice the effective exponent, which has been empirically found to be $\lambda_{\text{eff}} = (\lams+\laml)/4$. The long-dashed line is proportional to $e^{2 \lambda_{\rm eff} t}$. Note that this regime is only observable in the classical OTOC. Finally, the last line shows a strongly mixed case where the second crossover begins to be noticeable in the quantum curve.

To shed more light on these findings, we investigate a different indicator, namely the trace of the stability matrix. It is a fundamental one, in the sense that it directly probes the dynamics of the system without any choice of operators which could introduce artifacts \cite{Tabor_1989}. Additionally, it is invariant under canonical transformation, like the computation of the maximal Lyapunov exponent of the system. The main issue is that this probe does not have an equivalent in quantum mechanics, in which case we will resort to the OTOC. Nonetheless, both are linked as they probe the same quantity, i.e. the exponent that dictates the dynamic of the system.\\
\indent The trace of the stability matrix is defined as
\begin{align}
    \Tr(M(t)) &= \Tr 
    \begin{pmatrix*}
        \frac{\partial q(t)}{\partial q_0}     & \frac{\partial q(t)}{\partial p_0}      \\
        \frac{\partial p(t)}{\partial q_0}     & \frac{\partial p(t)}{\partial p_0}        \\
    \end{pmatrix*}\\
    &= \pm 2 \cosh(\lambda t)
\end{align}
with $q$ and $p$ the position and momentum of the system and $q_0$ and $p_0$ the initial position and momentum \cite{Haake_2001}. The last equality holds provided that the system is chaotic. The exponent dictating the dynamics can be extracted by
\begin{align}
        \lambda &=\frac{1}{t} \text{arccosh}\left(\frac{\abs{\Tr(M(t))}}{2}\right) \\
	&\simeq \frac{1}{t} \log\abs{\Tr(M(t))}. \label{eq:exponentFromTr}
\end{align}
It can be obtained numerically by sampling over a coherent state initially centred around the hyperbolic fixed point.

\section{Driven pendulum} \label{sec_analyticalDescription}
% =============================================================================================
\indent As stated in the introduction, we now focus our attention to the study of a periodically driven pendulum whose Hamiltonian is given by 
\begin{align} \label{eq_hamiltonian}
    H(q,p,t) = \frac{p^2}{2} + \lams^2 \cos(q) + \kappa \omega \lams \cos(q) \cos(\omega t+ \varphi)
\end{align}
with the first two terms being the regular pendulum Hamiltonian, and where $\kappa \omega \lams$ is the amplitude of the perturbation, $\omega$ its frequency, and $\lams$ the stability exponent associated with the hyperbolic fixed point at $(q,p)=(0,0)$ of the Hamiltonian. The dimensionless parameter $\kappa$ characterises the strength of the perturbation. As we assumed a weak driving amplitude with a high frequency, the system must satisfy  $\kappa \ll \lams \ll \omega $ and $\kappa\omega \sim 1$. This choice stems from the fact that the Hamiltonian \eqref{eq_hamiltonian} proves to be the one that universally describes the dynamics of a system in the vicinity a resonance \cite{Ozorio_de_Almeida_1984, Brodier_Schlagheck_Ullmo_2002}. More details can be found in \cite{Lichtenberg_Lieberman_2013} as well as Appendix \ref{app_secular} for for the two-degree-of-freedom case.\\
% We define $\kappa = K/\lams\omega \ll 1$ the perturbation parameter, assumed small. \\
\indent The equation of motion derived from the Hamiltonian \eqref{eq_hamiltonian} is
\begin{align} \label{eq_motion1}
    \ddot{q}(t) = \lams^2 \sin(q(t)) + \kappa \omega \lams \sin(q(t)) \cos(\omega t+\varphi)\,,
\end{align}
which is nonintegrable. A strategy to obtain an analytical solution is to separate the phase space into two regions. In the first one, labelled the \textit{linear region}, the equations of motion can be linearised around the origin, and in the second one, the \textit{homoclinic region}, the system only slightly deviates from the homoclinic. The crossover position between the two regions is labelled $\qcross$ and must be chosen as to satisfy both hypotheses: it must be small as compared to 1 in order to be able to linearise the dynamics around the origin, and at the same time it must be large enough so that the system only slightly deviate from the separatrix in the rest of the phase space. 
%
%
%
%
%%%%% Linear region %%%%%%%
%%%%%%%%%%%%%%%%%%%%%%%%%%%
\subsection{Linear region}
As the dynamics can be linearised around the origin, equation \eqref{eq_motion1} is approximated by 
\begin{align} \label{eq_motion_linearised}
    \ddot{q}(t) = \lams^2 q(t) + \kappa \omega \lams \cos(\omega t+\varphi)  q(t).
\end{align}
To solve this differential equation, we take its Laplace transform and obtain
\begin{align} \label{eq_qtilde}
  \tilde{q}(z) &= \frac{b(z)}{z^2-\lams^2} + \kappa \omega \lams  \frac{e^{i \varphi} \tilde{q}(z-i \omega) +e^{-i \varphi}\tilde{q}(z+i \omega)}{2(z^2-\lams^2)}
\end{align}
where $b(z) = z q_0 + p_0$ was introduced for conciseness and where $q_0$ qnd $p_0$ are the initial conditions when the system enters the linear region. The next step consists in injecting the expression of $\tilde{q}$ inside the terms of the right-hand side evaluated at $z\pm i \omega$. Doing so introduces corrections of order $\mathcal{O}(1/\omega^\alpha)$ with $\alpha=1,2$. We only consider corrections up to $\mathcal{O}(1/\omega)$ for the final answer. As the momentum is obtained by time derivating the position, a $\omega$ prefactor might appear and thus we need to keep the terms $\mathcal{O}(1/\omega^2)$ until the end. Doing so leads to
\begin{align} \label{eq_qtilde2}
    \begin{aligned} 
       \tilde{q}(z) = &\frac{b(z)}{z^2-\lams^2}+  \frac{\kappa \omega \lams e^{i \varphi} }{2(z^2-\lams^2)}\frac{b(z-i \omega)}{(z-i\omega)^2-\lams^2}\\
       &+\frac{\kappa \omega \lams e^{-i \varphi} }{2(z^2-\lams^2)}\frac{b(z+i \omega)}{(z+i\omega)^2-\lams^2}+ \mathcal{O}\left(\frac{1}{\omega^3}\right)
       \end{aligned}
\end{align}
The expression of $q$ is then obtained by taking the inverse Laplace transform of \eqref{eq_qtilde2}. In addition, we consider that the time is the one elapsed in the linear region and the phase is the initial phase plus the one accumulated during the whole motion. This means doing the shifts $t \to t-t_0$ and $\varphi \to \varphi + \omega t_0$, with $t_0$ the time elapsed before the system entered the linear region. The position in the linear region $q_{\rm lin}$ in leading order in $1/\omega$ is: 
\begin{align}
    q(t) = &q_0 \cosh(\lams (t-t_0)) \nonumber\\
    &+ \left( \frac{p_0}{\lams} - \kappa  \sin({\varphi}+ \omega t_0) q_0\right) \sinh(\lams (t-t_0)).
\end{align}
The momentum in the linear region is given by 
\begin{align}
    p_{\text{lin}}(t) = &\left(q_0 \lams +\kappa p_0 \sin(\omega t + \varphi) \right)\sinh(\lams (t-t_0)) \nonumber\\
    &+\left( p_0 + \kappa \lams q_0 (\sin(\omega t+\varphi)-\sin({\varphi}+\omega t_0))\right) \nonumber\\
    &\ \ \ \cross\cosh(\lams (t-t_0))
\end{align}
This approximation is valid at least up to the crossover position $q_{\text{c}} \ll 1$.\\
\indent The duration of the linear region can be obtained exactly by looking at the time it takes the system to reach $\pm q_{\text{c}}$, where the $\pm$ indicates whether the system is propagating towards the positive or negative position: 
\begin{align} \label{eq:durationlinear}
    t_{1} &= \frac{1}{\lams} \log \abs{ \frac{q_c + \sqrt{q_c^2 - q_0^2 + \left(\frac{p_0}{\lams} - \kappa\ q_0 \sin(\varphi)\right)^2}}{q_0 \left(1 - \kappa  \sin(\varphi) \right) + \frac{p_0}{\lams} }}.
\end{align}
After that time has elapsed, the system enters the homoclinic region.\\
\indent The stability matrix in this region is given by
\begin{align}
    M(t) &= 
    \begin{pmatrix*}
        \frac{\partial q_{\text{lin}}(t)}{\partial q_0}     & \frac{\partial q_{\text{lin}}(t)}{\partial p_0}      \\
        \frac{\partial p_{\text{lin}}(t)}{\partial q_0}     & \frac{\partial p_{\text{lin}}(t)}{\partial p_0}        \\
    \end{pmatrix*} \label{eq_Mlin}
\end{align}
with
\begin{align}
     \frac{\partial q_{\text{lin}}(t)}{\partial q_0} &=\cosh(\lams (t-t_0))  \nonumber\\
     & \ \ \ \ \ - \kappa  \sin({\varphi}+ \omega t_0) \sinh(\lams (t-t_0)),\\
     \frac{\partial q_{\text{lin}}(t)}{\partial p_0} &=\frac{1}{\lams} \sinh(\lams (t-t_0)), \\
     \frac{\partial p_{\text{lin}}(t)}{\partial q_0}  &= \lams \sinh(\lams (t-t_0)) \nonumber\\
     &\mkern-60mu + \kappa  (\sin(\omega t + \varphi) -\sin({\varphi}+ \omega t_0)) \lams \cosh(\lams (t-t_0)), \\
     \frac{\partial p_{\text{lin}}(t)}{\partial p_0} &= \cosh(\lams (t-t_0))\nonumber\\
     & \ \ \ \ + \kappa \sin(\omega t + \varphi) \sinh(\lams (t-t_0)).
\end{align}
The symplectic evolution is conserved as the determinant is $1$, up to some correction in $\kappa^2$ which are neglected in this work. Details on this derivation can be found in Appendix \ref{app_linear}.

%%%% Homoclinic region %%%%
%%%%%%%%%%%%%%%%%%%%%%%%%%%
\subsection{Homoclinic region}
When a trajectory propagates sufficiently far away from the linearised region around the unstable fixed point, the trajectory stays close to the homoclinic and its dynamics can be approximated by expanding the trajectory in terms of the homoclinic motion. An ansatz for the solution is then given by
\begin{align}
    q_{\rm h }(t) =q_{\text{sep}}(t) + \delta q(t),
\end{align}
with $q_{\rm sep}$ the position of the separatrix and $\delta q$ the deviation form it such that $\abs{\delta q} \ll q_{\rm sep}$.\\
\indent We first derive the expression of the separatrix from \eqref{eq_motion1} with $\kappa=0$, i.e. the unperturbed homoclinic in the absence of a driving. The solution is
\begin{align} \label{eq_homoclinic}
    q_{\text{sep}}(t) &= 4 \arctan\left[ \tan(\frac{q_1}{4}) \exp\left[\text{sgn}(q_1)\lams(t-t_1)\right] \right]
\end{align}
where we define $\text{sgn}(q_1)= q_1/|q_1|$ which indicates again whether the system evolves towards the positive (outward motion) or negative positions (inward motion), $q_1= \pm q_{\text{c}}$ is the position when the system enters the homoclinic region, and $t$ is the time elapsed in it. Details are found in appendix \ref{app_homoclinic}.\\
\indent Let us now obtain an expression for the deviation from the separatrix $\delta q$ by considering again the presence of the driving and assuming initial conditions $(q_1, p_1)$ not necessarily on the separatrix. Multiplying equation \eqref{eq_motion1} by $\dot{q}$ and integrating with respect to time yields a first integral of motion
\begin{eqnarray}
\frac{1}{2} \dot{q}^2(t) & \simeq & \delta E_1 + 2 \lams^2 \sin^2 [q(t)/2] \nonumber \\ && + \kappa \lams \sin(\omega t + \varphi) \dot{q}(t) + O(\kappa/\omega)  \label{eq:qdot2}
\end{eqnarray}
with $q_1 = q(t_1)$, $p_1 = \dot{q}(t_1)$, and
\begin{eqnarray}
\mkern-20mu \delta E_1 & = & \frac{p_1^2}{2} - \lams^2 ( 1 - \cos q_1) - \kappa \lambda_s \sin(\omega t_1 + \varphi) p_1 \sin q_1 \nonumber \\
& \simeq & \frac{1}{2}(p_1^2 - \lams^2 q_1^2 ) - \kappa \lams \sin(\omega t_1 + \varphi) p_1 q_1 + O(q_1^3)
\end{eqnarray}
the deviation, $\delta E_1 = E_1 - E_{\text{sep}}$, of the initial energy $E_1 = H(q_1,p_1,t_1)$ from the unperturbed separatrix energy $E_{\text{sep}} = \lams^2$.
Using the generic scalings $\kappa \ll 1$ and $\kappa \omega \sim \lams$ resulting from secular perturbation theory (see Appendix \ref{app_secular}), terms of the order $\sim \kappa / \omega$ effectively scale as $\sim \kappa^2 / \lams \sim \lams/\omega^2$ and are henceforth neglected.\\
\indent Consider now a tiny deviation from the above separatrix solution. The energy deviation from the separatrix is small as well, and one can assume $\delta E_1 \ll \lams^2 \sin^2[q(t)/2]$ for all $t$, provided $q(t)$ starts sufficiently far away from the unstable fixed point and does not approach too closely the fixpoint on the other side of the separatrix arc.
Taking the positive square root of Eq.~\eqref{eq:qdot2} and performing a Taylor series expansion in $\delta E_1 /\{\lams^2 \sin^2[q(t)/2]\}$ yields
\begin{eqnarray}
\dot{q}(t) & \simeq & 2 \lams \sin[q(t)/2] + \kappa \lams \sin(\omega t + \varphi) \sin q(t) \nonumber \\ && +\frac{\delta E_1}{2 \lams \sin[q(t)/2]} - \frac{\delta E_1^2}{16 \lams^3 \sin^3[q(t)/2]} \,, \label{eq:qdot}
\end{eqnarray}
where we neglect terms $\sim O(\delta E_1^3)$ that will turn out to scale quadractically either with $\kappa$ or with the initial phase-space deviation from the separatrix.\\
\indent Injecting our assumption that the position in the homoclinic region is a small deviation from the homoclinic yields a differential equation for $\delta q$:
\begin{eqnarray}
\delta \dot{q}(t) & \simeq & \left( - \lams \tanh \tau(t) + \frac{\delta E_1}{8 \lams} \sinh[2\tau(t)]\right) \delta q(t) \nonumber \\ && + \text{sgn}(q_1)\left(\frac{\delta E_1}{2 \lams} \cosh \tau(t)  - \frac{\delta E_1^2}{16 \lams^3} \cosh^3 \tau(t)\right) \nonumber \\ && - \kappa \lams \sin(\omega t + \varphi) \left( 2 \,\text{sgn}(q_1) \frac{\sinh \tau(t)}{\cosh^2\tau(t)} \right. \nonumber \\ && \left. - \frac{\sinh^2 \tau(t) - 1}{\cosh^2 \tau(t)} \delta q(t) \right) \,, \label{eq:dqdot}
\end{eqnarray}
neglecting terms that scale as $O(\kappa^2)$, $O(\kappa/\omega)$, $O(\delta E_1^3)$, $O(\delta E_1^2 \delta q)$, and $O(\delta q^2)$, and where $\tau(t) = \lams(t-t_1) + \log(\tan(q_1/4))$.
The solution to this differential equation is given by  
\begin{eqnarray}
\frac{\delta q(t)}{\text{sgn}(q_1)} & = & \frac{\delta E_1 \lams(t - t_1)}{4 \lams^2\cosh\tau(t)} \left( 1 + \frac{\delta E_1}{16 \lams^2}\left[2 \cosh^2\tau(t) - 3\right]\right) \nonumber \\ && + \frac{\delta E_1\left\{ \sinh[2\tau(t)] - \sinh[2\tau(t_1)]\right\}}{8 \lams^2\cosh\tau(t)} \nonumber \\ && \ \ \ \times \left( 1 + \frac{\delta E_1}{8 \lams^2} \left[ \cosh^2 \tau(t) - 2 \right] \right) \nonumber \\ && - \frac{\delta E_1^2\left\{\sinh[4\tau(t)] - \sinh[4\tau(t_1)] \right\}}{256 \lams^4 \cosh \tau(t)} .\label{eq_epsAnalytical}
\end{eqnarray}
This approximation is valid as long as the dynamics can be linearised around the homoclinic. \\
\indent Again, we can obtain the duration of the region, which we will note $t_2$. As $\delta q$ contributes little to that duration, it can be safely neglected. Assuming the homoclinic region starts at $q_1$ and ends at $q_2$, one has
\begin{align}\label{eq_durationhomoclinic}
    t_{2} &= \frac{1}{\lams}  \log\left(\frac{\tan(\frac{q_2}{4})}{\tan(\frac{q_1}{4})}\right) \simeq \frac{1}{\lams} \log\left( \frac{16}{q_{\rm c}^2} \right).
\end{align}
where $q_2= 2\pi - q_{\text{c}}$ or $q_{\text{c}}$ depending if the system propagates towards to positive or negative position respectively. It is important to note that injecting $t_2$ in $q_{\text{sep}}+\delta q$ will not return exactly $q_2$ because the deviation was not considered. This is however not an issue as initial conditions for the linear region will still fulfill the linearisation hypothesis and thus any deviations will be taken into account in the coming linear region.\\
\indent The associated stability matrix can be calculated by evaluating the effect of a tiny initial deviation with respect to this particular reference trajectory, at fixed total propagation time $t_2$:
\begin{equation} \label{eq_Msep}
M = \left( \begin{array}{cc} \frac{\delta q_2}{\delta q_1} & \frac{\delta q_2}{\delta p_1} \\ \frac{\delta p_2}{\delta q_1} & \frac{\delta p_2}{\delta p_1}\end{array} \right) \simeq \left( \begin{array}{cc} 0 & (1 - \kappa_1)/\lams \\ - (1 + \kappa_1)\lambda_\text{s} & \kappa_2 - \kappa_1 \end{array} \right)
\end{equation}
with $\kappa_j = \kappa \sin(\omega t_j + \varphi)$, $j=1,2$. The symplectic structure is conserved as well. A more detailed derivation is shown in Appendix \ref{app_homoclinic}. \\
\indent As we split the motion such that our mathematical framework allows us to know the final position and momentum of one region using the final conditions of the previous region as initial ones, we refer to our development as a \textit{mapping}.

%%%%%% Scaling of q_c %%%%%
%%%%%%%%%%%%%%%%%%%%%%%%%%%
\subsection{Scaling of the crossover position with the perturbation}
\begin{figure}
    \begin{center}
    	\includegraphics[width=0.4\textwidth, angle=0, trim={0cm 0cm 0cm 0cm}, clip]{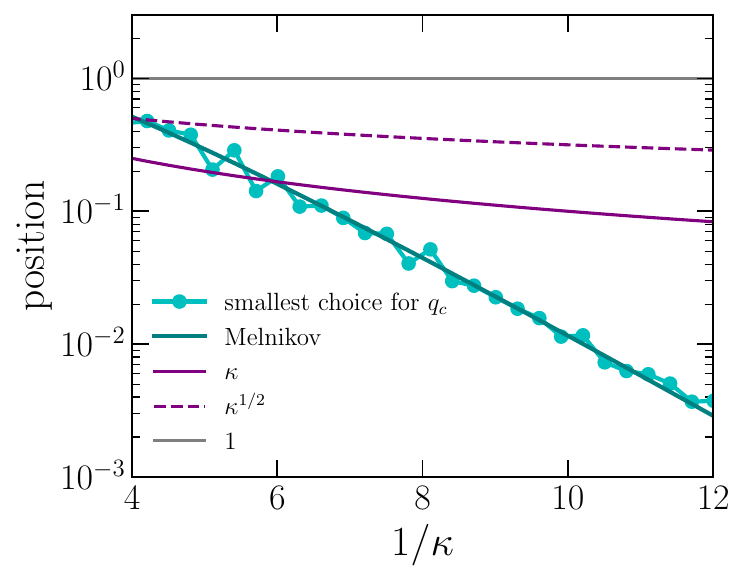}
    \end{center}
    	% \captionsetup{justification=justified}%,margin=2cm}
    \caption{Closest position to the origin reachable before the homoclinic region formalism deviates from the numerical solution, plotted as a function of the inverse perturbation parameter $1/\kappa = \omega \lams/K$, obtained numerically (blue) and with the Melnikov method detailed in Appendix \ref{app_sec_melnikov} (darker shade of blue). The points were obtained using, $\kappa \omega = \lams$, $\varphi=0.1$. In solid purple, we show the closest position to the origin set by the linearisation condition of equation \eqref{eq:qdot}, and in dashed purple an arbitrary scaling of $q_{\rm c} = \kappa^\alpha$, with $\alpha=1/2$, that satisfies both boundaries.}
    \label{fig_scaling_qc}
\end{figure}
In this section, we analyse how $\qcross$ can be chosen while still fulfilling the conditions of validity of the linear and homoclinic regions as a function of the perturbation. The upper limit of the choice of $\qcross$ is set by the former: as the equation should be linearisable, $\qcross \ll 1$. The lower limit is set by the latter. As the system evolves, the hypothesis of small deviation from the homoclinic is eventually not satisfied. Indeed, as a phase-space trajectory gets closer to the origin, its momentum remains finite whereas the momentum of the homoclinic tends to 0. Identifying where the quasi-integrable case deviate from the integrable one enables us to know where this hypothesis breaks down puts a lower bound on the choice of $q_{\text{c}}$. To this end, two criteria must be investigated. \\
\indent Firstly, it can be shown that a trajectory evolved with the perturbed Hamiltonian will remain close to one that is propagated using the unperturbed Hamiltonian for a time that increases exponentially as the perturbation decreases \cite{Nekhoroshev_1977,Poschel_1993}. We can use the Melnikov method \cite{Arnold_1964, Melnikov_1962, Chirikov_1979, Lichtenberg_Lieberman_2013} to estimate the scaling of the lower bound of the crossover position. As shown in Appendix \ref{app_sec_melnikov}, we obtain that the choice of $q_c$ must satisfy 
\begin{equation} \label{eq_melnikovBound}
q_c \gg  \sqrt{\frac{\sqrt{2}\pi \kappa (\omega/\lambda_\text{s})^3}{\sinh(\pi \omega/2 \lambda_\text{s})}},
\end{equation}
letting us a generous freedom of choice for $q_\text{c}$. 
We then verify numerically those predictions. To this end, we investigate how close, from initial conditions taken near the hyperbolic fixed point, the system was able to get to the other fixed point linked to the first one by a separatrix before the exact numerical solution deviated from our analytical expression as a function of $\kappa$. We consider two solutions not to match anymore when the relative difference of momentum, estimated by $\abs{p_1 - p_2}/\left( \abs{p_1}+\abs{p_2} \right)$, becomes larger than a given threshold \footnote{The threshold was set at 0.5, but its value is of little importance since any threshold will yield an exponential scaling.}. The distance between the position of deviation and the hyperbolic fixed point determines the lower bound of $q_c$. Those numerical verifications match perfectly with the lower bound \eqref{eq_melnikovBound} obtained using the Melnikov method. None\-theless, this lower bound is not the only one to take into account.\\
\indent Indeed, one must also ensure that all the steps that were followed to obtain the homoclinic region formalism are valid. From equation \eqref{eq:qdot}, it implies
\begin{align}
    2\delta E_1 + 2\kappa \lams \sin(\omega t +\varphi)\dot{q}(t) \ll 4 \lams^2 \sin^2\left(\frac{q(t)}{2}\right).
\end{align}
Using again \eqref{eq:qdot} for the expression of $\dot{q}$ and neglecting terms of order $\mathcal{O}(\kappa^2)$, i.e. keeping only the term related to the separatrix, the inequality becomes 
\begin{align}
    \kappa \left(q_{\rm c}^2 + \sin(\omega t+\varphi)\sin\left(\frac{q(t)}{2}\right) \right) \ll \sin^2\left(\frac{q(t)}{2}\right).
\end{align}
This equation can become invalid if $q_{\rm c}$ is chosen too small, leading to a disagreement at the end of the homoclinic region between the equations before and after linearisation around the separatrix, i.e. \eqref{eq:qdot2} and \eqref{eq:qdot}. From the equation one infers the condition $\kappa \ll q_{\rm c}$, which puts a lower bound on the scaling of the boundary position.\\
\indent These results are shown in figure \ref{fig_scaling_qc}, which shows the minimal value for the choice of $q_c$ as a function of the inverse of the perturbation parameter. The numerical bound is shown in blue and is indeed exponential in the inverse of $\kappa$ \cite{Nekhoroshev_1977,Poschel_1993}. The bound related to Nekoroshev estimates matches perfectly with the numerical one. The smaller the perturbation parameter, the closer to the origin the boundary $q_c$ can be set. However, the bound related to the linearisation condition, in solid purple, proves to be more limiting, and is therefore the one to follow for small $\kappa$. Let us however note that for a larger perturbation parameter, the limiting condition would be the bound obtained using the Melnikov method, but as it is outside the scope of our work, we do not consider it here.\\
\indent As a summary, as long as $q_c$ is chosen as larger than the purple line while still being smaller than 1, represented by the grey horizontal line, i.e. $\kappa \ll q_{\rm c} \ll 1$, the linear-homoclinic formalism will work. A working choice is $q_{\text{c}}=\kappa^\alpha$, with for instance $\alpha = 1/2$.

%%%%%%%%%%%%%%%%%%%%%%%%%%%%%%%%%%%%%%%%
%%%% Trace of the stability matrix %%%%%
%%%%%%%%%%%%%%%%%%%%%%%%%%%%%%%%%%%%%%%%
\section{Exponential growth rate from the trace of the stability matrix} \label{sec_expGrowthFromTr}
\begin{figure}
	%\centerline{
    \includegraphics[width=0.475\textwidth, angle=0, trim={0cm 0cm 0cm 0cm}, clip]{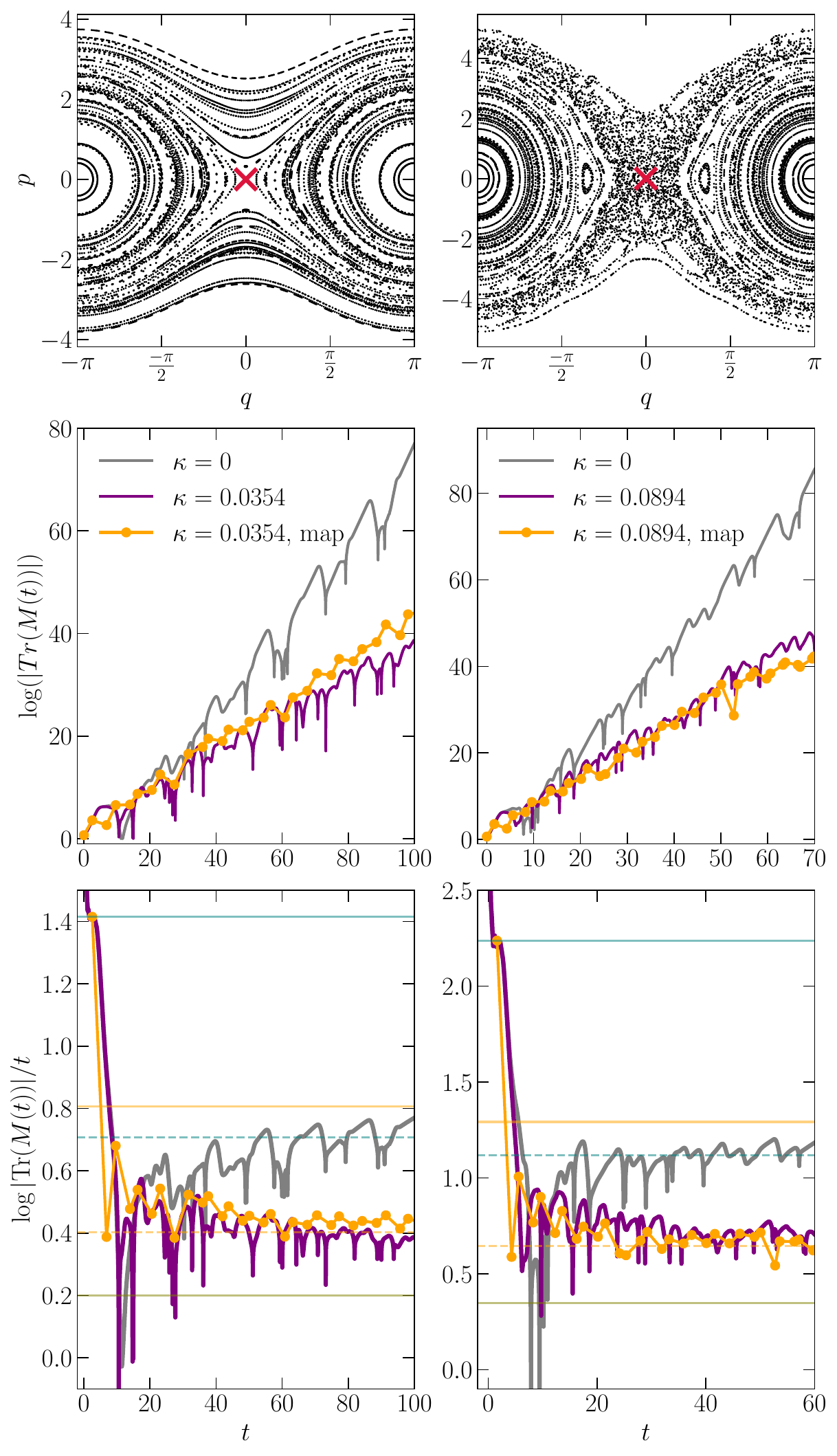}
	%}
	% \captionsetup{justification=justified}%,margin=2cm}
	\caption{\textit{First row}: stroboscopic section of the driven pendulum. \textit{Second row}: trace of the stability matrix for different perturbation parameters. \textit{Third row}: Exponential growth rate extracted from the second row. Different rates are show as horizontal lines: $\lams$ in blue, $\lams/2$ in dashed blue, $(\lambda_{\text{L}}+\lams)/2$ in orange, and $(\lambda_{\text{L}}+\lams)/4$ in dashed orange. The last two rows use the same colour scheme: grey for the numerical solution of the simple pendulum, purple for the driven pendulum, and orange for the driven pendulum using the mapping derived in \ref{sec_analyticalDescription}. The first column is for $\kappa=0.0354$, the second one for $\kappa=0.0894$}
    \label{fig_stroboscopicPSandTr_2param}
\end{figure}
\indent In Section \ref{sec_analyticalDescription}, we developed a mathematical formalism describing the system by splitting the phase space into two regions. This means in particular that the trace of the stability matrix can be obtained using this mapping. Bearing in mind that when splitting the motion in different regions, the corresponding stability matrices must be multiplied {\color{red} ref}, one can write 
    \begin{align}
        \Tr(M(t)) &= \Tr(... M_{\rm h}(t_{k+1}) M_{\text{lin}}(t_k) ...).
    \end{align}
\indent Stroboscopic sections for a near-integrable and a mixed regime are shown on the first line of figure \ref{fig_stroboscopicPSandTr_2param}. The hyperbolic fixed point is located at the origin of the axes at $(0,0)$. On the second row of the same figure we show the absolute value of the trace of the stability matrix obtained from the numerical solution of \eqref{eq_motion1} for the simple pendulum (grey), for the driven pendulum numerically (purple) and using our mapping (orange) for a single point in the vicinity of the separatrix. On the last row, we show the exponential growth rate extracted from the second row using the same colour coding as well as horizontal lines corresponding to multiples of the stability exponent, the Lyapunov exponent of the chaotic sea, and the effective exponent.\\
%
%
%% 1-cycle approximation %%
%%%%%%%%%%%%%%%%%%%%%%%%%%%
\indent First, we notice that the trace in the absence of driving follows the same behaviour as highlighted in \cite{Steinhuber_Schlagheck_Urbina_Richter_2023}, i.e. an initial exponential growth where the rate is $\lams$, then $\lams/2$. In the presence of the perturbation, the long time behaviour is different and the rate lies between the Lyapunov exponent and the stability exponent. \\
\indent We can see that the mapping allows to reproduce faithfully the trace of the stability matrix. Consequently, it enables us to extract analytically an exponent. As one must propagate the system to long time in order to obtain a reliable expression of the exponent dictating the dynamics, this means multiplying a high number of stability matrices, which of course quickly becomes untractable. Nonetheless, one can try to obtain a reasonably good estimate of the exponent by considering only a total of 2 linear and one homoclinic regions and by taking the initial conditions at the negative-position boundary of the linear region, on the homoclinic. The reason behind this choice is the following. As the goal is to obtain an estimate of the effective exponent, a choice would be to consider a cycle, i.e. one linear and one homoclinic region. However, doing so results in the particular case where the trace of the stability matrix vanishes. To counter that, we considered an additional region. One could have made the choice to consider two homoclinic regions and one linear region. While this would have slightly modified the analytical result, the estimation would still agree with the numerical simulations.\\
\indent The estimate of the trace in this region is
    \begin{align}
	\Tr(M(t_1+t_2+t_1)) &= \Tr(M_{\text{lin}}(t_1)M_{\rm h}(t_2) M_{\text{lin}}(t_1)) \label{eq:TrDeriv} 
\end{align}
where $(q_0, p_0) = (-q_c, \lams q_c+\delta p)$, with $\delta p\ll \lams q_{\rm c}$, are the initial conditions of the linear region, $(q_1, p_1)$ those of the homoclinic region, $t_1$ and $t_2$ the durations of the linear and homoclinic regions. For these initial conditions, the times are given by 
\begin{align}
    \begin{aligned} \label{eq_t1t2forlambda}
        t_1 &= \frac{1}{\lams} \log \left(\frac{2 q_{\rm c}}{\delta p + \kappa q_{\rm c} \sin(\varphi)} +1 \right) 
        \simeq \frac{1}{\lams} \log \left(\frac{2 }{\kappa}+1 \right) \\
        t_2 &= \frac{1}{\lams} \log\left(\frac{16}{q_{\rm c}^2}\right)
    \end{aligned}
\end{align}
where the expression of $t_1$ holds because it can be shown that the order of magnitude of $\delta p$ is $\kappa q_{\rm c} \lams$.\\
\indent Extracting the exponent yields
    \begin{align} \label{eq_lambda_expression1}
	\lambda_{\rm eff} = \frac{1}{2t_1+t_2} \log \abs{\Tr(M_{\text{lin}}(t_1) M_{\text{sep}}(t_2) M_{\text{lin}}(t_1))}.
    \end{align}
Using the stability matrices \eqref{eq_Mlin} and \eqref{eq_Msep}, the asymptotic behaviour of the hyperbolic functions $\cosh(x) \sim \sinh(x) \sim \exp(t)/2$, as well as the durations in \eqref{eq_t1t2forlambda}, the dominant contribution to the trace of the stability matrix is given by
\begin{align}
   &\abs{\Tr(M_{\text{lin}}(t_1) M_{\text{sep}}(t_2) M_{\text{lin}}(t_1))} \nonumber\\
   &= \abs{\frac{8\kappa}{q_{\rm c}^2} \left(\sin(\varphi + \omega t_2) - 2\sin(\varphi +\omega t_1) - \sin(\varphi)\right)} + \mathcal{O}\left(\kappa^2\right) \\
   &\simeq \frac{ \beta \kappa}{q_{\rm c}^2} + \mathcal{O}\left(\kappa^2\right)
\end{align}
with $\beta>0$ some real coefficient and $\kappa$ assuming without loss of generality to be positive. The combination of sine functions in the parenthesis is akin to a random prefactor. Thus, one can replace it with the square root of the average of the square of all terms, which yields an estimation of $\beta = 16\sqrt{2} \simeq 20$\\
\indent Taking this into account as well as the scaling of $q_{\rm c}$ as a power of $\kappa$, one obtains the following expression for the exponent dictating the dynamics of the system:
\begin{align}
    \lambda_{\rm eff} &= \lams\frac{\log\left(\beta \kappa^{1-2\alpha}\right)}{\log\left(  16/\kappa^{2+2\alpha}\right)}.
\end{align}
A suitable choice of $\alpha$ that ensures the hierarchy $\kappa \ll q_{\rm c} \ll 1$ is $\alpha = 1/2$, which results in the following expression for the effective exponent:
\begin{align} \label{eq:lambda2}
    \lambda_{\rm eff} &= \lams\frac{\log\left(\beta \right)}{\log\left(  16/\kappa^{3}\right)}.
\end{align}

For completeness, let us also present the expression that would result from two homoclinic propagations between which one propagation in the linear region takes place. 
The exponent has then the form
\begin{align} \label{eq:lambdaalt}
    \lambda_{\rm eff} &= \lams\frac{\log\left(2/\kappa \right)}{\log\left(  16^2/\kappa^{2+4\alpha}\right)}
\end{align}
which, using $\alpha=1/2$, becomes
\begin{align} \label{eq:lambda2alt}
    \lambda_{\rm eff} &= \lams\frac{\log\left(2/\kappa \right)}{\log\left(  16^2/\kappa^{4}\right)}.
\end{align}

\begin{figure}[H]
	\centerline{
		\includegraphics[width=0.35\textwidth, angle=0, trim={0cm 0cm 0cm 0cm}, clip]{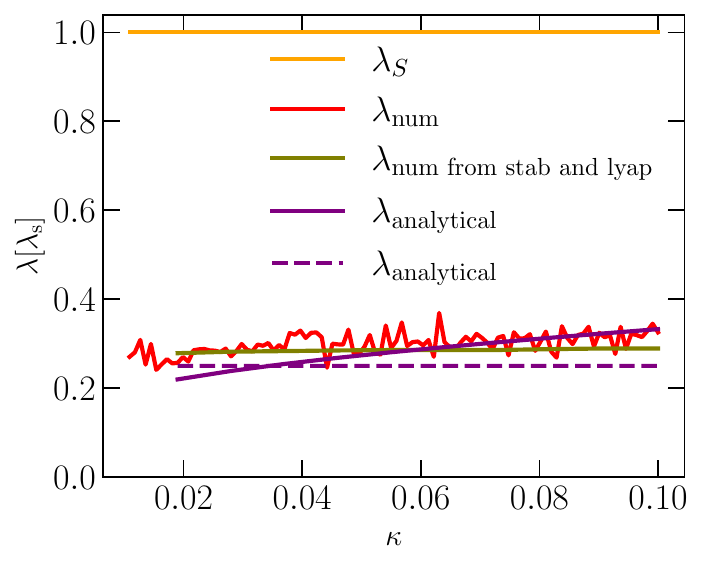}
	}
	% \captionsetup{justification=justified}%,margin=2cm}
	\caption{Exponential growth rates. The stability exponent $\lams$ is shown in orange, the effective exponent of the trace $\lambda_{\text{num}} = (\lams+\lambda{_\text{L}})/4$, obtained numerically is shown in red, $\lambda_{\text{num}}$ but with both exponents obtained using the Wolf algorithm in green, equation \eqref{eq:lambda2}, labelled $\lambda_{\text{analytical}}$ is shown in purple using $\beta=20$, and \eqref{eq:lambda2alt} in dashed purple, using $q_{\text{c}}=\kappa^{1/2}$ for the last two.}
    \label{fig_effectivelambda_numericsvsPrediction}
\end{figure}

These expressions can be compared with the numerics, as is shown on figure \ref{fig_effectivelambda_numericsvsPrediction}. In this figure, we show different exponents in unit of the stability exponent: the stability one $\lams$ (orange), $(\lams+\lambda_{\text{L}})/4$ where the two exponents have been obtained separately using the Wolf algorithm \cite{Wolf_Swift_Swinney_Vastano_1985} (green), $(\lams+\lambda_{\text{L}})/4$ obtained from the evolution of the trace of the stability matrix (red), and finally the analytical expressions \eqref{eq:lambda2} (purple), using $\beta=20$ and \eqref{eq:lambda2alt} (dashed purple). The predictions are consistent with what is observed numerically. 
%
%
%
%
%%%%%%%%%%%%%%%%%%%%%%%%%%%%%%%%
%%% Mixed systems %%%%%%%%%%%%%%
%%%%%%%%%%%%%%%%%%%%%%%%%%%%%%%%
\section{Complementary numerical findings}
\label{sec_compl}
\subsection{Out-of-time-ordered correlators} \label{sec_otherSystOTOC}
The exact application of expression \eqref{eq:lambda2} to other systems is not straightforward. However, as the Hamiltonian of the driven pendulum is the one that universally describes any system around a resonance, the results of the previous section can be extended to any system in the near integrable regime. We focus on the kicked rotor and the kicked Bose-Hubbard dimer. Furthermore, to link the quantum and classical dynamics, we use the OTOC as a probe.\\
\begin{figure}
	\centerline{
		\includegraphics[width=0.475\textwidth, angle=0, trim={0cm 0cm 0cm 0cm}, clip]{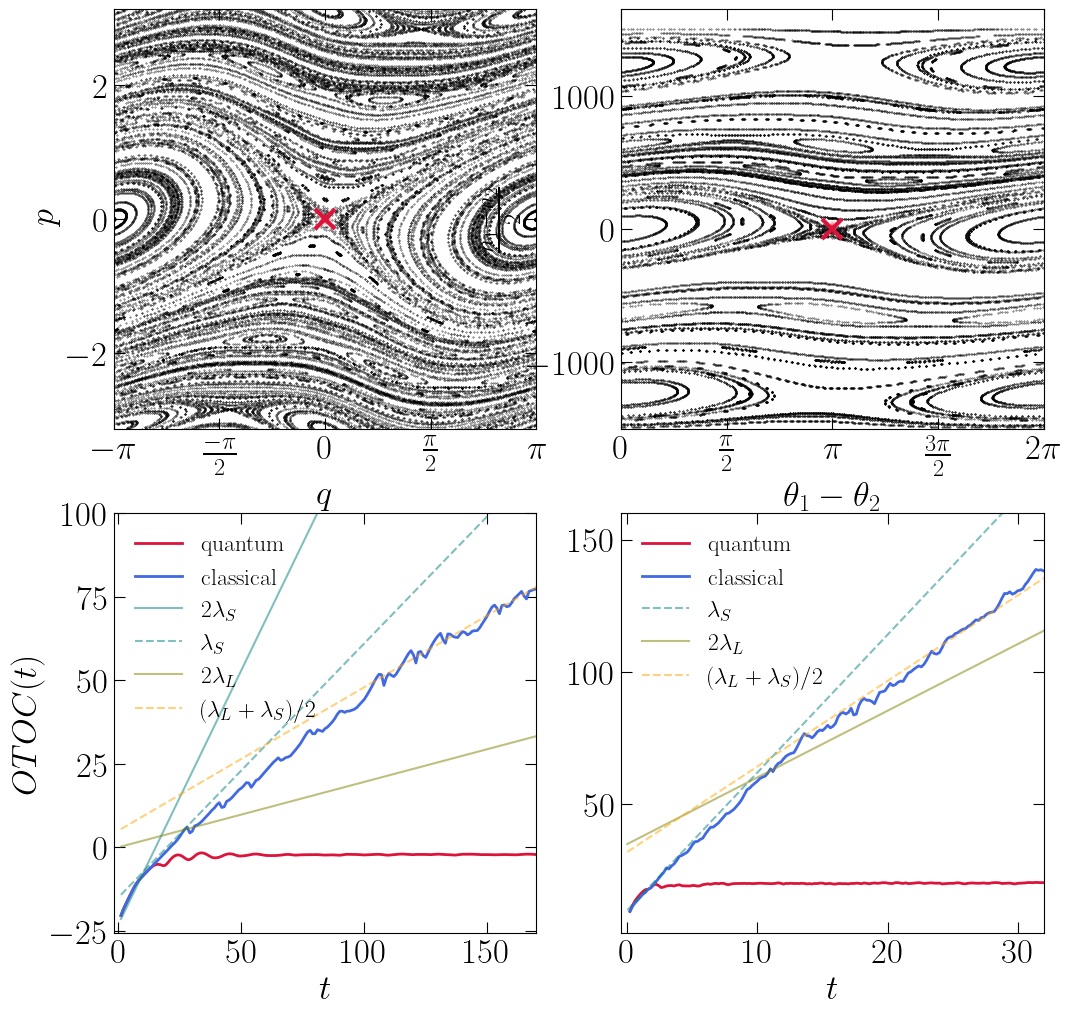}
	}
	% \captionsetup{justification=justified}%,margin=2cm}
	\caption{\textit{First colum}: Standard map/kicked rotor. Top: Phase space representation of the standard map for $K=0.6$, $\hbar_{\text{eff}} = 2^{-14}$. The initial coherent state is shown in red, centred at $(0,0)$. 
    Bottom: quantum (red) and classical (blue) OTOC with the operator $\hat{A} = \hat{B} = \hat{p}$. Different exponential function are shown to highlight the different regimes. 
    \textit{Second column}: kicked dimer. Top: Phase space representation of the kicked dimer for $N=3000$, $U=0.025J$. The initial coherent state is shown in red.
    Bottom: quantum (red) and classical (blue) OTOC with the operator $\hat{A} = \hat{n}_1$, $\hat{B} = \hat{n}_2$. Different exponential function are shown to highlight the different regimes.}
    \label{fig_KRandKD}
\end{figure}
\indent The kicked rotator is defined by the Hamiltonian 
\begin{align} \label{eq_kickedRotor}
    \hat{H}(t) = \frac{\hat{p}^2}{2} + K \cos\hat{\theta} \sum_{n=-\infty}^\infty
\delta\left(t-n\right)
\end{align}
where $K$ is the kicking strength \cite{Rozenbaum_Ganeshan_Galitski_2017}. Its classical limit, knows as the standard map, can be written
\begin{align}
\begin{aligned}
    p_{n+1} &= p_n + K \sin\theta_n\\
    \theta_{n+1} & = \theta_n + p_{n+1}.
\end{aligned}
\end{align}
In the first column of figure \ref{fig_KRandKD} we show in the top panel a phase space representation for a kicking %strength $K=0.6$,
and in the bottom panel the OTOC with the operator $\hat{A} = \hat{B} = \hat{p}$ alongside different exponentials to highlight the different regimes of exponential growth rate. First, the OTOC grows with an exponential rate of $2\lambda_s$, transitions to a $1 \lambda_s$ regime as expected \cite{Steinhuber_Schlagheck_Urbina_Richter_2023}, and then reaches the $(\lambda_s+\lambda_L)/4$ regime. We thus observe the same behaviour as with the trace of the stability matrix, the only difference being the factor two in the growth rate due to the modulus square appearing in \eqref{OTOC_def}.
Another system that we can study is the kicked dimer \cite{Strzys_Graefe_Korsch_2008, Liang_Zhang_Chen_2024}. It can be seen as a Bose-Hubbard dimer where the hopping is activated by time-discrete kicks. Its Hamiltonian is 
\begin{align} \label{eq_kickedDimer}
    \hat{H} = &\frac{U}{2} \left( \hat{n}_1 \left(\hat{n}_1 -1\right) + \hat{n}_2 \left(\hat{n}_2 -1\right) \right) \nonumber\\
    & - J \left (\hat{a}^\dagger_1 \hat{a}_2 + \hat{a}^\dagger_2 \hat{a}_1 \right) \tau \sum_{n=-\infty}^\infty \delta \left( t - n \tau \right)
\end{align}
where $U$ is the on-site interaction parameters, $\hat{n}_l = \hat{a}^\dagger_l \hat{a}_l$ the population operator on site $l$, and $\hat{a}^\dagger_l$ and $\hat{a}_l$ the creation and annihilation operators on site l, and $\tau$ the period and strength of the hopping kicks. The classical Hamiltonian can be obtained by taking its Weyl symbol and is
\begin{eqnarray}
    H(\bm{q}, \bm{p}, t) &= &\frac{U}{2} \left( n_1(n_1-1) + n_2(n_2-1) \right) \nonumber \\
    &&- J \left( q_1 q_2 + p_1 p_2 \right) \tau \sum_{n=-\infty}^\infty \delta \left( t - n \tau \right)
\end{eqnarray}
with $n_l = (q_l^2 + p_l^2 - 1)/2$.\\
\indent Similarly to the kicked rotor, we show a stroboscopic section in the first line of the second column of \ref{fig_KRandKD}, followed by the OTOC with $\hat{A} = \hat{n}_1$, $\hat{B} = \hat{n}_2$ the population operators on site $1$ and $2$. \\

\subsection{Mixed systems and exponent hierarchies} \label{sec_mixedAndHierarchies}

\begin{figure}
	\centerline{
		\includegraphics[width=0.475\textwidth, angle=0, trim={0cm 0cm 0cm 0cm}, clip]{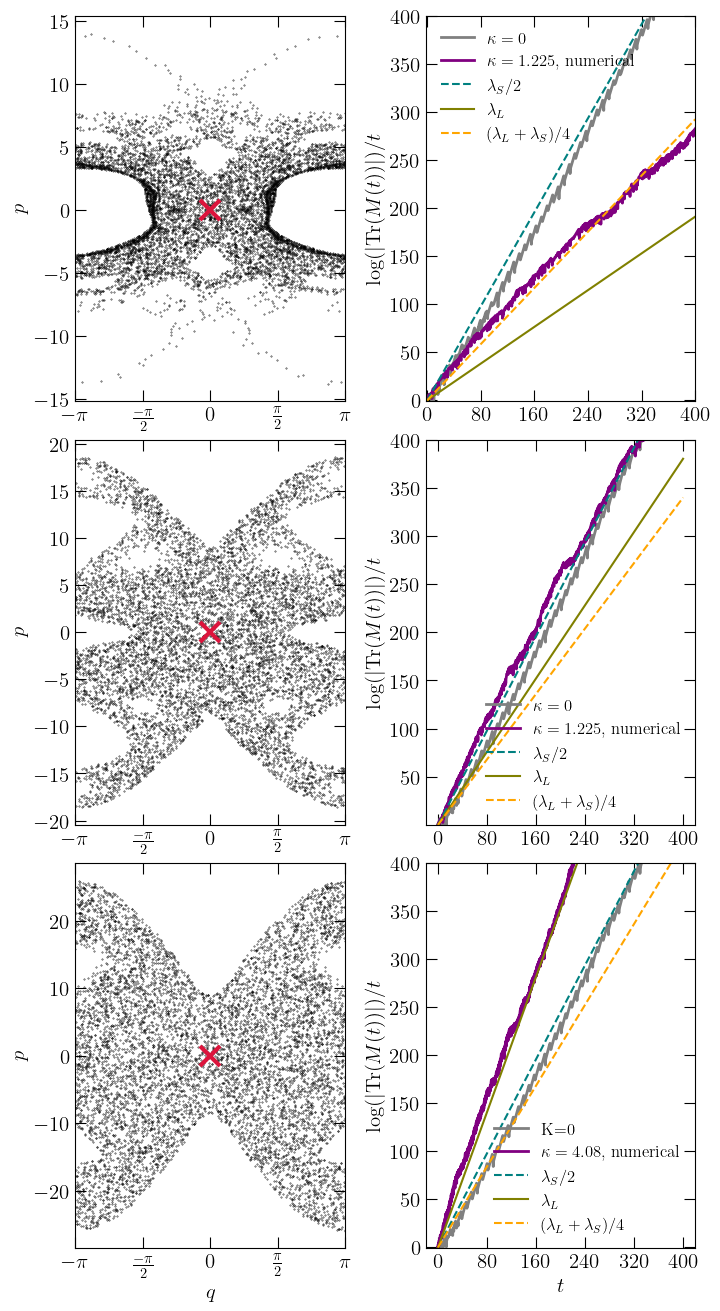}
    }
	\caption{\textit{First column}: Phase space of the driven pendulum for $\lams=\sqrt{6}$ and different perturbation parameters. \textit{Second column}: trace of the stability matrix in the corresponding phase space for a initial coherent state centred on the fixed point (orange cross in the phase space). Lines corresponding to exponentials with the rate being the half the stability exponent $\lambda_s/2$, the Lyapunov exponent $\lambda_L$, or a combination of both are shown. For all: $\lambda_s=\sqrt{6}$.  First row: $\kappa=0.245$, $\lambda_L = 0.479$. Second row: $\kappa=1.225$, $\lambda_L = 0.951$. Third row: $\kappa = 4.08$, $\lambda_L = 1.76$. }
    \label{fig_mixed_pendulum}
\end{figure} 
Our formalism is built on the hypothesis on a thin chaotic layer around the homoclinic for a driven pendulum. It is however useful to see how it fares against a larger one, i.e. well beyond the hypothesis of a small perturbation $\kappa$. The results are shown in figure \ref{fig_mixed_pendulum}, using again the trace of the stability matrix. The phase space is represented in the first column, and the trace of the stability matrix in the second one. As the phase space becomes more chaotic, the hierarchy of the different exponent changes. In the first row of \ref{fig_mixed_pendulum}, it is  $\lambda_L < (\lambda_s+\lambda_L)/4 < \lambda_s/2$. In the second row, the two smallest exponent are switched: $(\lambda_s+\lambda_L)/4 < \lambda_L < \lambda_s/2$. Finally, in the third row, we have $(\lambda_s+\lambda_L)/4 < \lambda_s/2 < \lambda_L$. We can notice the different behaviours depending on the hierarchy. In the first one, the long-time exponent (before going to the Lyapunov regime at very long times) is $(\lambda_s+\lambda_L)/4$. In the second, $\lambda_s$ dominates, and in the last one, it is the Lyapunov exponent $\lambda_L$.\\
\begin{figure}[t]
\begin{center}
	\includegraphics[width=0.45\textwidth, angle=0, trim={0cm 0cm 0cm 0cm}, clip]{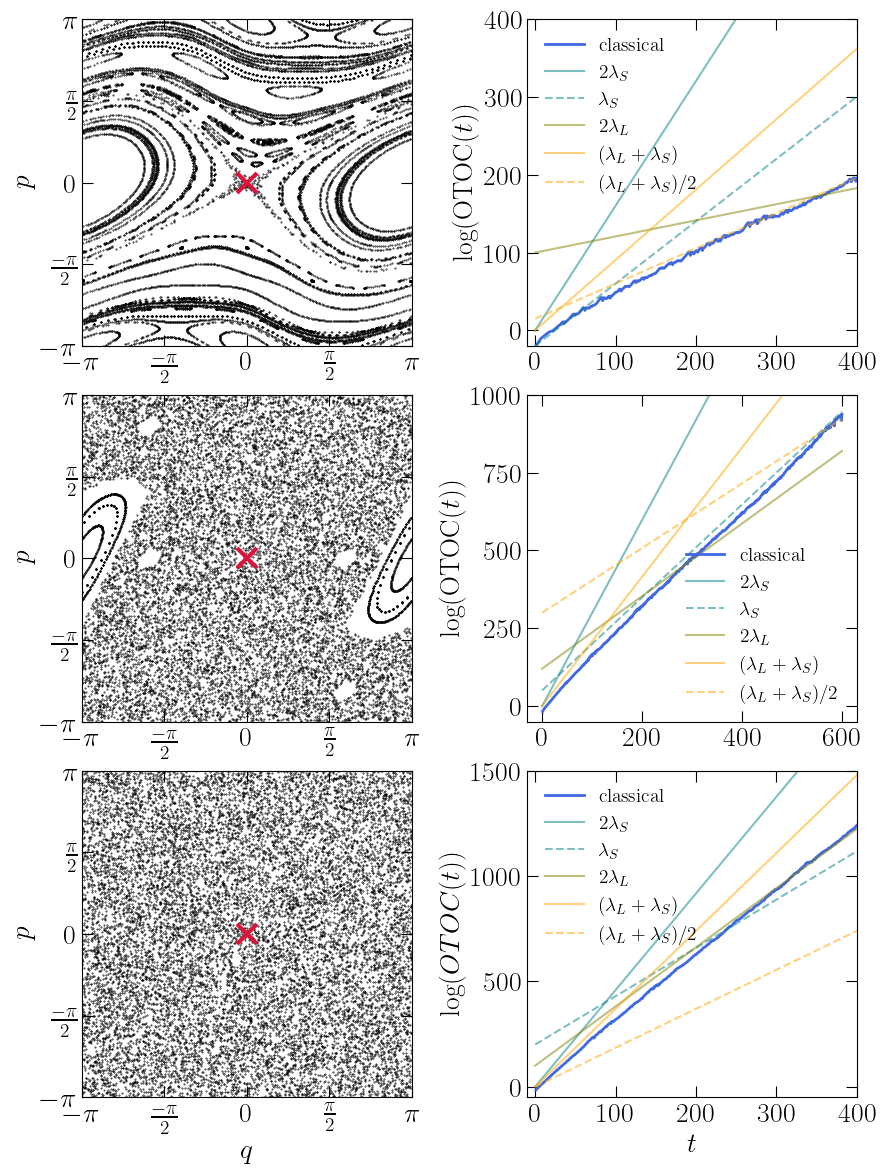}
\end{center}
%\captionsetup{justification=justified}%,margin=2cm}
\caption{Phase space (first column) and OTOC (right column) for different kicking parameters of the kicked rotor. The right plot also shows lines corresponding to exponentials with the rate being the stability exponent $\lambda_s$, the Lyapunov exponent $\lambda_L$, or a combination of both. Second row: $K=0.676$, $\lambda_s=0.802$, $\lambda_L = 0.104$. Second row: $K=3.007$, $\lambda_s=1.49$, $\lambda_L = 0.583$. Third row: $K=8.00$, $\lambda_s=2.29$, $\lambda_L = 1.40$. The operators for the OTOC are $\hat{A} = \hat{B} = \sin(\hat{p})$.}
    \label{fig_mixed_KR}
\end{figure} 
\indent We also investigated at the different hierarchies in the kicked rotor. We show in the first column of figure \ref{fig_mixed_KR} different phase spaces of the kicked rotor corresponding to different hierarchy of $\lams$, $\laml$ and $(\lams+\laml)/4$. The systems are strongly mixed or fully chaotic. In the right column, we show out-of-time-ordered correlators for an initial coherent state centred on the hyperbolic fixed point at the origin, and we highlight the different exponential regimes with straight lines. The same conclusions as for the driven pendulum are drawn.
%
%
%
%
%

%%%%%%%%%%%%%%%%%%%%%%%%%%%%%%%%
%%% Long time %%%%%%%%%%%%%%
%%%%%%%%%%%%%%%%%%%%%%%%%%%%%%%%
\subsection{Long-time behaviour} \label{sec_longTime}

\begin{figure}
	\centerline{
		\includegraphics[width=0.5\textwidth, angle=0, trim={0cm 0cm 0cm 0cm}, clip]{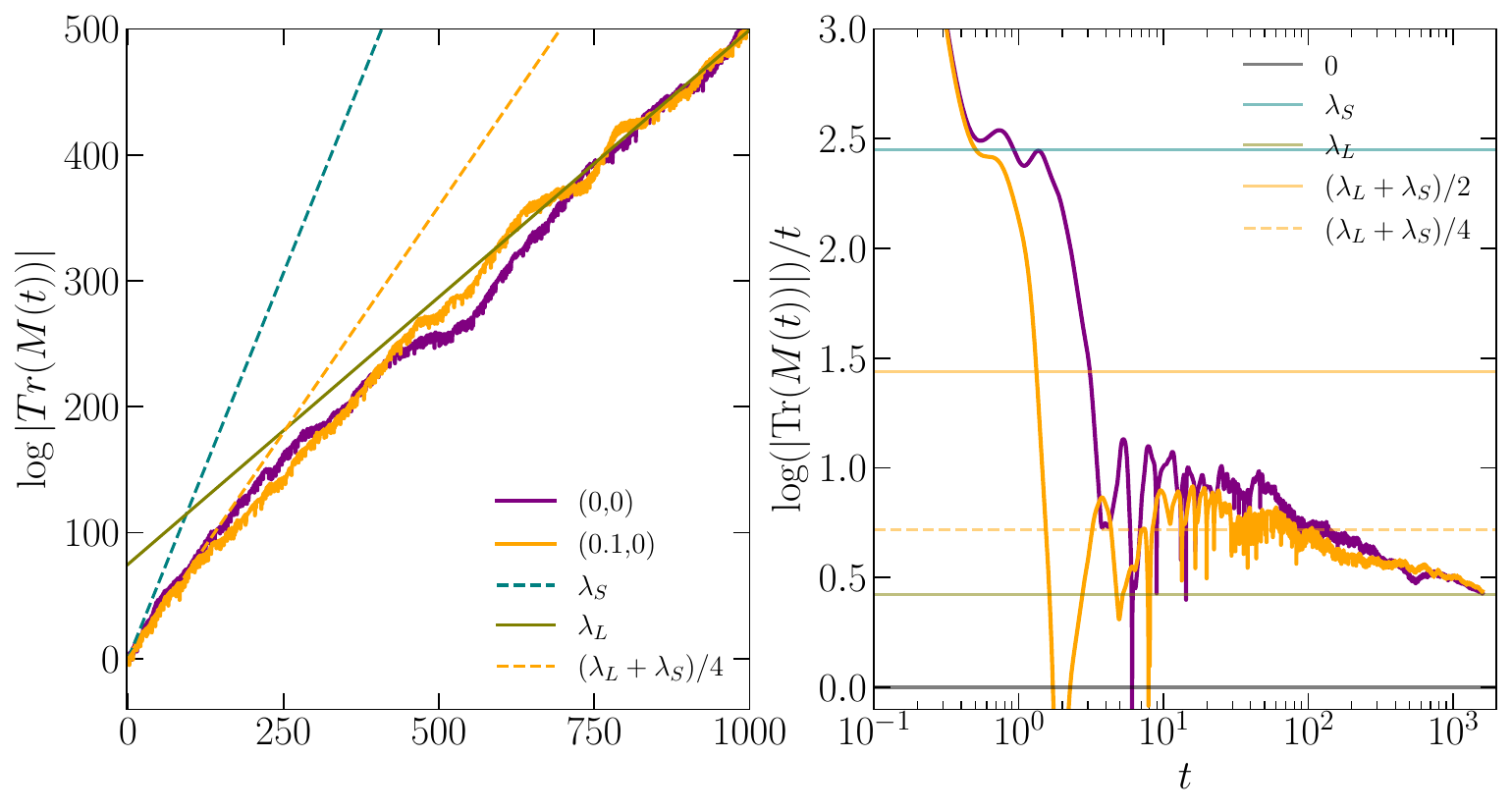}
	}
	\caption{\textit{Top}: trace of the stability matrix of the driven pendulum for $V=6$, $K=3.007$, $\omega=10$, $\hbar_{\text{eff}}= 2^{-8}$, for a coherent state centred on $(0,0)$ (purple) and $(0.2, 0.2)$ (orange). \textit{Bottom}: Exponential growth rate. Note the horizontal log scale. }
    \label{fig_longTime}
\end{figure} 
Whether the system is integrable or not, the initial $\lams$ regime of the trace as well as the contribution of $\lams$ to the later-time exponent are purely features of the initial wavepacket being placed on the hyperbolic fixed point. One can then wonder what happens at very long time, and how the behaviour varies with the thickness of the chaotic layer. On figure \ref{fig_longTime} we show the exponential growth rate of the trace of the stability matrix for an initial coherent state centred either on the hyperbolic fixed point or in the chaotic layer around the homoclinic in the driven pendulum. The trace first follows a $\lams$ regime, which is shorter the farther from the hyperbolic fixed point the initial state is. It then goes to a $(\lambda_s + \lambda_L)/4$ regime, the system eventually falls back on $\lambda_L$ regime. The farther from the fixed point the initial state is, the shorter the $\lams$ regime. However, it does not seem to have a large impact on the long time behaviour, indicating that being on the separatrix is more important than close to the fixed point.

\section{Conclusion}
% =============================================================================================

In summary, our study revealed an intricate regular-to-chaotic transition of exponential growth laws of scrambling signatures that are associated with unstable fixed points of the underlying classical dynamics.
While in the purely integrable and fully chaotic cases the growth laws of OTOCs and related observables are exclusively governed by the (local) instability and (global) Lyapunov exponents, $\lams$ and $\laml$, respectively, (with characteristic crossovers $2\lams \to\lams$ occurring in the integrable case \cite{Steinhuber_Schlagheck_Urbina_Richter_2023}), effective exponents interpolating between these two limiting cases emerge in generic near-integrable and mixed regular-chaotic systems.
In the near-integrable regime in particular, a crossover to an OTOC growth law with the exponent $(\lams + \laml)/2$ is found for long evolution times in the deep semiclassical limit.
This finding is corroborated by an analytical study of a more elementary (classical) scrambling signature, namely the mean trace of the stability matrix, in the framework of the rapidly driven pendulum system which can be seen as a standard model describing the emergence of chaos near nonlinear resonances.
Scaling estimates of the effective exponent with the perturbation from integrability can thus be yielded, in good agreement with numerical simulations.

Most of the results shown here, including those related to OTOC growth behaviors, are essentially classical. 
One can thus wonder to what extent the intricate growth crossovers that we found for near-integrable dynamics are actually observable in quantum systems, given the smallness of the Planck cell that would be required in order to avoid saturation before reaching the crossover time scale.
This investigation is left for future studies.
We also postpone to future investigations a more involved analysis of scrambling signatures that are associated with unstable fixed points in a strongly mixed regular-chaotic context, possibly giving, or not, rise to scar states \cite{Heller} in the corresponding quantum system.
This could yield interesting tools to probe the existence of 
such localized states \cite{Sinha_Ray_Sinha_2021} which could be particularly valuable in the context of complex quantum systems \cite{HummelRichterSchlagheck2023}.

\acknowledgements
MS acknowledges funding through the Studienstiftung des Deutschen Volkes.

\appendix

\section{Secular perturbation theory}
\label{app_secular}

In this appendix, we review the basic calculation steps of secular classical perturbation theory, with the aim to motivate the choice of the driven pendulum Hamiltonian \eqref{eq_hamiltonian}.
More details on this perturbative approach to solve the classical equations of motion near a resonance can be found in the standard literature on nonlinear Hamiltonian dynamics \cite{Lichtenberg_Lieberman_2013}.

Consider a two-degrees-of-freedom (2-dof) system described by two approximate action variables $I_1,I_2$ and the associated angle variables $\theta_1,\theta_2$.
Near integrability, the Hamiltonian describing this system is written as
\begin{equation}
H(I_1,I_2,\theta_1,\theta_2) = H_0(I_1,I_2) + H_1(I_1,I_2,\theta_1,\theta_2) \label{app_eq:H0H1}
\end{equation}
with $H_0$ the unperturbed part and $H_1$ a weak perturbation.
The latter can be represented as a $2\pi$-periodic function in the angle variables,
\begin{equation}
H_1(I_1,I_2,\theta_1,\theta_2) = \sum_{k_1,k_2=-\infty}^\infty V_{k_1,k_2}(I_1,I_2)e^{i(k_1\theta_1 + k_2\theta_2)} \,, \label{app_eq:H1Fourier}
\end{equation}
with the Fourier coefficients $V_{k_1,k_2} = V_{-k_1,-k_2}^*$.
While no specific assumption is made for the former, we shall be particularly interested in using this 2-dof autonomous system in order to emulate a 1-dof system under perturbative periodic driving, described by the Hamiltonian $H_0^{(1)}(I_1) + H_1(I_1,\theta_1,\omega t)$ with $\omega$ the driving frequency.
For this purpose one can choose
\begin{equation}
H_0(I_1,I_2) = H_0^{(1)}(I_1) + \omega I_2 \label{app_eq:Hdriving}
\end{equation}
and identify $\omega I_2$ with the negative (time-dependent)  energy of the driven 1-dof system, through considering the submanifold of vanishing total energy $H = 0$ within the 2-dof system's phase space.

For a sufficiently weak deviation from integrability, the angle variable dependence of the Hamiltonian $H$ can be eliminated through canonical perturbation theory \cite{Lichtenberg_Lieberman_2013}, provided the system is sufficiently far away from important resonances of its unperturbed dynamics.
This is the essence of the Kolmogorov-Arnold-Moser (KAM) theorem.
Classical motion thus continues to evolve along invariant tori in phase space, as in the strictly integrable case.
Near a resonance, however, the canonical perturbative approach fails to converge and the invariant tori break up. 
New substructures, reminiscent of pendulum dynamics, emerge at their place in the phase space, as predicted by the Poincaré-Birkhoff theorem.

Consider the occurrence of such a $r$:$s$ resonance in action space, where $r\in\mathbb{N}$ full cycles of the angle variable $\theta_2$ are exactly matched by $s\in\mathbb{N}$ full cycles of $\theta_1$.
At the corresponding action variables $I_1 = I_1^{r:s}$ and $I_2 = I_2^{r:s}$ we thus have
\begin{equation}
r\Omega_1(I_1^{r:s},I_2^{r:s}) = s\Omega_2(I_1^{r:s},I_2^{r:s}) \,, \label{app_eq:resonance}
\end{equation}
with
\begin{equation}
\Omega_j(I_1,I_2) = \frac{\partial H_0}{\partial I_j}(I_1,I_2)
\end{equation}
the unperturbed angular frequencies associated with the degrees of freedom $j=1,2$.
To describe the motion in the vicinity of this resonance, we first perform a canonical transformation to a frame that corotates with one of the degrees of freedom.
This transformation introduces new pairs of canonically conjugated phase-space variables $(J,\vartheta)$ and $(p,q)$,
\begin{equation}
    (I_1,\theta_1,I_2,\theta_2) \stackrel{F}{\longmapsto} (J,\vartheta,p,q) \,,
\end{equation}
and is defined via the generating function
\begin{equation}
F(J,p,\theta_1,\theta_2) = (I_1^{r:s} + p) \theta_1 + \left( I_2^{r:s} + \frac{1}{r}(J - s p) \right)\theta_2
\end{equation}
that depends on the new (approximate) action variables $J,p$ and the old (approximate) angle variables $\theta_1,\theta_2$.
The old (approximate) action and new (approximate) angle variables are then obtained according to
\begin{eqnarray}
I_1 = \frac{\partial F}{\partial\theta_1} & = & I_1^{r:s} + p \,,\\
I_2 = \frac{\partial F}{\partial\theta_2} & = & I_2^{r:s} +  \frac{1}{r}(J - s p) \,, \\
q = \frac{\partial F}{\partial p} & = & \theta_1 - \frac{s}{r} \theta_2 \,, \\
\vartheta = \frac{\partial F}{\partial J} & = & \frac{1}{r} \theta_2 \,.
\end{eqnarray}
The classical Hamiltonian \eqref{app_eq:H0H1} is then (exactly) rewritten in terms of these phase-space variables as
\begin{equation}
H(I_1,I_2,\theta_1,\theta_2) = \tilde{H}_0(J,p) + \tilde{H}_1(J,p,\vartheta,q)
\end{equation}
with
\begin{eqnarray}
\tilde{H}_0(J,p) & = & H_0[I_1^{r:s}+ p,I_2^{r:s} + (J-s p)/r] \,, \\
\tilde{H}_1(J,p,\vartheta,q) & = & H_1[I_1^{r:s}+ p,I_2^{r:s} + (J-s p)/r, \nonumber \\ && q + s \vartheta, r \vartheta] \,. \label{app_eq:H1corot}
\end{eqnarray}

The transformation to this corotating frame is particularly meaningful in the vicinity of the $r$:$s$ resonance where one of the new angle variables, namely $q$, will evolve much more slowly with time than the other one.
This can be seen from Taylor expanding $\tilde{H}_0$ in the vicinity of the resonance, given by $J=p=0$.
Owing to the resonance condition \eqref{app_eq:resonance} one has $\partial \tilde{H}_0/\partial p|_{J=p=0} = 0$ whereas the complementary derivative $\partial \tilde{H}_0/\partial J|_{J=p=0} = \Omega_2(I_1^{r:s},I_2^{r:s}) / r$ yields a finite frequency.
Our interest are the deviations of the action variables in leading order from the resonance, thus we obtain
\begin{equation}
\tilde{H}_0(J,p) \simeq E_{r:s} + \omega_{r:s} J + \frac{p^2}{2 m_{r:s}} +O(J^2, Jp, p^3) \label{app_eq:Happrox}
\end{equation}
for the unperturbed Hamiltonian, with the resonance parameters
\begin{eqnarray}
E_{r:s} & = & H_0(I_1^{r:s},I_2^{r:s}) \,, \\
\omega_{r:s} & = & \frac{1}{r} \Omega_2(I_1^{r:s},I_2^{r:s}) \,, \\
\frac{1}{m_{r:s}} & = & \left( \frac{\partial \Omega_1}{\partial I_1} - 2 \frac{s}{r} \frac{\partial \Omega_1}{\partial I_2} + \frac{s^2}{r^2} \frac{\partial \Omega_2}{\partial I_2}\right)(I_1^{r:s},I_2^{r:s})
\end{eqnarray}
(noting that $\partial \Omega_1/\partial I_2 = \partial \Omega_2 / \partial I_1$ for a twice continuously differentiable $H_0$).
Note that we deliberately neglect cross-terms $\propto Jp$ in the above expression \eqref{app_eq:Happrox}, to simplify the discussion.
These cross-terms translate the fact that the resonance condition \eqref{app_eq:resonance}, which traces a curve in action space, generally shifts with varying $J$.
For the particularly interesting case of a periodically driven system, for which we would have the unperturbed Hamiltonian \eqref{app_eq:Hdriving}, those cross-terms vanish and the expression \eqref{app_eq:Happrox} is exact up to corrections scaling as $O(p^3)$.

Employing the Fourier series representation \eqref{app_eq:H1Fourier}, the perturbation term \eqref{app_eq:H1corot} is in leading order written as
\begin{eqnarray}
V_{r:s}(q,\vartheta) & = &  H_1(I_1^{r:s},I_2^{r:s}, q + s \vartheta, r \vartheta) \nonumber \\ & = & \sum_{k_1,k_2=-\infty}^\infty V_{k_1,k_2}^{r:s} e^{i k_1 q} e^{i (s k_1 + r k_2) \vartheta} \label{app_eq:VFourier}
\end{eqnarray}
with $V_{k_1,k_2}^{r:s} = V_{k_1,k_2}(I_1^{r:s},I_2^{r:s})$.
The resulting total Hamiltonian $\tilde{H}_0(J,p) + V_{r:s}(q,\vartheta)$ generates a trivial time evolution for $\vartheta$, through the associated Hamilton equation $\dot{\vartheta} = \partial \tilde{H}_0/\partial J = \omega_{r:s}$.
This entitles us to make the substitution  $\vartheta = \vartheta_0 + \omega_{r:s} t$ in Eq.~\eqref{app_eq:VFourier} and thus describe the system in terms of the equivalent driven 1-dof Hamiltonian
\begin{equation}
H_{r:s}(p,q,t) = \frac{p^2}{2 m_{r:s}} + V_{r:s}(q,\omega_{r:s}t) \,. \label{app_eq:Hrs}
\end{equation}

Adiabatic perturbation theory \cite{Lichtenberg_Lieberman_2013}, a variant of canonical perturbation theory, allows one to eliminate the dependence on the rapidly evolving angle variable $\vartheta$ through an infinitesimal canonical transformation of the phase-space coordinates, thus yielding an integrable system.
In practice, this procedure amounts to replacing the perturbation term \eqref{app_eq:VFourier} by its $\vartheta$-average over one period,
\begin{equation}
\bar{V}_{r:s}(q) = \frac{1}{2\pi}\int_0^{2\pi} V_{r:s}(q,\vartheta) d\vartheta = \sum_{l=-\infty}^\infty V_{lr, -ls}^{r:s} e^{i l r q} \,.
\end{equation}
Neglecting the constant $V_{0,0}^{r:s}$ and using the fact that low-order Fourier coefficients generally dominate over high-order components, we can limit the above sum to the terms $l = \pm 1$.
With $\bar{V}^{r:s} \equiv |V_{r,-s}^{r:s}| = |V_{-r,s}^{r:s}|$ this yields the effective pendulum Hamiltonian
\begin{equation}
\bar{H}_{r:s}(p,q) = \frac{p^2}{2 m_{r:s}} + 2 \bar{V}^{r:s} \cos[r(q - q_0)] \label{app_eq:Hpend}
\end{equation}
for some offset angle $q_0$, featuring $r$ elliptic and $r$ hyperbolic fixpoints as requested by the Poincaré-Birkhoff theorem.

Going beyond the adiabatic approximation, Eq.~\eqref{app_eq:Hrs} in combination with Eq.~\eqref{app_eq:VFourier} provides a meaningful setting to investigate universal characteristics related to the emergence of chaos near the hyperbolic fixpoints of the above pendulum Hamiltonian \eqref{app_eq:Hpend}.
To simplify things as much as possible, let us specifically focus on the most elementary case of a $1$:$1$ resonance, $r=s=1$.
Restricting ourselves to the most dominant nontrivial Fourier coefficients $V_{\pm1,0}^{1:1}$ in the expansion of $V_{1:1}$ allows us to rewrite Eq.~\eqref{app_eq:Hrs} in leading order as
\begin{eqnarray}
H_{1:1}(p,q,t) & \simeq & \frac{p^2}{2 m_{1:1}} + 2 \bar{V}^{1:1} \cos(q - q_0) \nonumber \\ && + 2 |V_{1,0}^{1:1}| \cos(q + \omega_{r:s}t + \varphi)
\end{eqnarray}
for some other offset angle $\varphi$.
Setting $q_0 = 0$ without loss of generality, the above expression is rewritten as
\begin{eqnarray}
H_{1:1}(p,q,t) & \simeq & \frac{p^2}{2 m_{1:1}} + 2 \bar{V}^{1:1} \cos(q) \nonumber \\ && +2 |V_{1,0}^{1:1}| \cos(q) \cos(\omega_{r:s}t + \varphi) \nonumber \\ && - 2 |V_{1,0}^{1:1}| \sin(q)\sin(\omega_{r:s}t+\varphi) \,.
\end{eqnarray}
This is essentially equivalent to the Hamiltonian \eqref{eq_hamiltonian} that we focus on in our study, with the only difference that the parity-breaking term $\propto \sin q$ is omitted in the latter.

\section{Solution in the linear region} \label{app_linear}
Let us obtain the equations of motion of the system in the linear region. The starting point is \eqref{eq_motion_linearised}, written again here for convenience:
\begin{align} \label{app_eq_motion_linearised}
    \ddot{q}(t) = \lambda_{\text{s}}^2 q(t) + \kappa \omega \lambda_{\text{s}} \cos(\omega t+\varphi)  q(t)
\end{align}
The first step consist in taking the Laplace transform of the equation in order to remove the time derivatives and only deal with an algebraic equation. \\
\indent The Laplace transform of a function $f$ is given by 
\begin{align}
        \tilde{f}(z) \equiv \mathcal{L}(f)(z) = \int_0^{+\infty}f(t) e^{- z t} dt
\end{align}
where both notations were introduced as one might be prefer over the other depending on the length of the term. While the Laplace transform of most terms are standard cases, the one containing the driving is slightly different and is given by
\begin{align}
    &\mathcal{L}(\kappa \lams \omega \cos(\omega t + \varphi) q(t))(z)\nonumber\\
    &= \frac{\kappa \lams \omega}{2}\left(e^{i \varphi} \tilde{q}(z-i \omega) + \frac{\kappa \lams \omega}{2}e^{-i \varphi} \tilde{q}(z+i \omega) \right).
\end{align}
Isolating in the new equation the Laplace transform of the position, $\tilde{q}$, one obtains 
\begin{align}
    \tilde{q}(z) &= \frac{b(z)}{z^2-\lams^2} + \kappa \omega \lams  \frac{e^{i \varphi} \tilde{q}(z-i \omega) +e^{-i \varphi}\tilde{q}(z+i \omega)}{2(z^2-\lams^2)}
\end{align}
where $b(z) = z q_0 + p_0$ was introduced for conciseness and where $q_0$ qnd $p_0$ are the initial conditions when the system enters the linear region.\\
\indent This equation is not directly exactly solvable. To obtain a solution, we inject the expression of $\tilde{q}$ evaluated at $z\pm i \omega$:
\begin{align}
\begin{aligned}
       &\tilde{q}(z) = \frac{b(z)}{z^2-\lams^2}\\
       &+ \kappa \omega \lams e^{i \varphi}  \frac{b(z-i \omega)+\frac{\kappa \omega \lams}{2} \left(e^{i \varphi}\tilde{q}(z-2i \omega)+  e^{-i \varphi} \tilde{q}(z)\right)}{2(z^2-\lams^2)((z-i\omega)^2-\lams^2)}\\
       &+ \kappa \omega \lams e^{-i \varphi} \frac{b(z+i \omega)+\frac{\kappa \omega \lams}{2}\left(e^{i \varphi}\tilde{q}(z)+e^{-i \varphi} \tilde{q}(z+2i \omega)\right)}{2(z^2-\lams^2)((z+i \omega)^2-\lams^2)}
       \end{aligned}
\end{align}
One could go on indefinitely by injecting again the expression of $\tilde{q}$ evaluated at $z\pm 2 i \omega$. However, one can notice that the dominating term is the first one, which scales as the inverse of the argument. This means that $\tilde{q}(z+ \alpha \omega) = \mathcal{O}((z+\alpha \omega)^{-1})$ since $b(z)\propto z$. The terms with the argument $z\pm 2 i \omega$ can thus be safely neglected as we restrict our development to $\mathcal{O}(1/\omega^2)$ included for now\footnote{We will only keep the first order for the position and the momentum, but as obtaining the latter is done by time derivating the former, this introduces some $\omega$ prefactor. To be consistent in our approximation, we must keep the $\mathcal{O}(1/\omega^2)$ and neglect them only after the derivation.}, using the fact that $\omega \gg \lams$ with $\kappa \omega \sim \lams$. We thus obtain
\begin{align} \label{app_eq_qtilde}
\begin{aligned} 
       \tilde{q}(z) &= \tilde{q}_{\rm I}(z)+\tilde{q}_{\rm II}(z)+\tilde{q}_{\rm II}(z)+\mathcal{O}\left(\frac{1}{\omega^3}\right)
       \end{aligned}
\end{align}
with
\begin{align}
    \tilde{q}_{\rm I}(z) &= \frac{b(z)}{z^2-\lams^2}\\
     \tilde{q}_{\rm II}(z) &= \frac{\kappa \omega \lams  }{2(z^2-\lams^2)} \left( \frac{e^{i \varphi} b(z-i \omega)}{(z-i\omega)^2-\lams^2}+\frac{e^{-i \varphi}b(z+i \omega)}{(z+i\omega)^2-\lams^2}\right)\\
      \tilde{q}_{\rm III}(z) &=\frac{\kappa^2 \omega^2 \lams^2 b(z)}{4(z^2-\lams^2)^2 }\left(\frac{1}{(z-i\omega)^2-\lams^2}+ \frac{1}{(z+i\omega)^2-\lams^2} \right)
\end{align}

\indent To obtain the position $q$, one must perform the inverse Laplace transform of \eqref{app_eq_qtilde}, which is given by 
\begin{align} \label{app_eq_inverseLaplace}
    q(t) = \frac{1}{2\pi i} \int_\Gamma \tilde{q}(z) e^{z t} \dd z
\end{align}
with $\Gamma: [-R, R]\to \mathbb{C}, x \mapsto \gamma+ i x$ in the limit $R \to \infty$, with $\text{Re}(\gamma)>\gamma_0$, with $\gamma_0$ the real part of the singularity of $\tilde{q}$ with the highest real part. As usual, we use the residue theorem to cpmpute this integral. The closed path we use is $\Gamma_\circ = \Gamma \circ\Gamma_R$, with $\Gamma_R: [\pi/2, 3\pi/2]\to \mathbb{C}, \theta \mapsto Re^{i \theta}$. In the limit $R\to \infty$, all the singularities of $\tilde{q}$ are encompassed by $\Gamma_\circ$ by construction, once, and counterclockwise. We can thus apply the residue theorem, which in this case reads
\begin{align} \label{app_eq_residueTheorem}
    \frac{1}{2\pi i}\int_{\Gamma} \tilde{q}(z) e^{z t} \dd z &= \sum_{z_i} \text{Res}(z_i)
\end{align}
with the sum being on all the singularities $z_i$ of the integrand of the left-hand side, $\text{Res}(z_i)$ the residue of the integrand in $z=z_i$, and where only the integral along $\Gamma$ remains in the limit $R\to \infty$, the one along $\Gamma_R$ tends to zero.
\\
\indent As the Laplace transform and its inverse are linear, let us consider the inverse transform of $\tilde{q}_{\rm I}$, $\tilde{q}_{\rm II}$ and $\tilde{q}_{\rm III}$ separately. 
The inverse transform $\tilde{q}_{\rm I}$, labelled $q_{\rm I}$, is given by
\begin{align}
    q_{\rm I}(t) = \cosh(\lams t) + \frac{p_0}{\lams} \sinh(\lams t)
\end{align}
and corresponds to the hyperbolic behaviour that the system would have even in the absence of driving.\\
\indent The inverse Laplace transform of $\tilde{q}_{\rm II}$, labelled $q_{\rm I}$, is 
\begin{align} 
\begin{aligned}
    &q_{\rm II}(t) = \frac{\kappa}{4\lams^2  + \omega^2} \biggl[ \lams \cosh(\lams t) \biggl( 2p_0 \sin(\omega t + \varphi)\\
    & - q_0 \omega \cos(\omega t + \varphi)+ \lams (q_0 \omega \cos(\varphi) - 2p_0 \sin(\varphi))\biggr) \\
    &+ \sinh(\lams t) \biggl( 2\lams^2 q_0  \sin(\omega t + \varphi)- p_0 \omega \cos(\omega t + \varphi) \\
    & -2\sin(\varphi) \lams^2 q_0 -\omega(p_0 \cos(\varphi) + q_0 \omega \sin(\varphi) \biggr) \biggr].
\end{aligned}
\end{align}
As $\kappa \ll \lams \ll \omega $ and $\kappa \omega \sim \lams$, only the very last term is of order $1/\omega$. However, the terms with the time dependence in trigonometric functions have to be kept for the momentum, as taking the time derivative introduces an additional $\omega$ prefactor, thus rendering the term of order $1/\omega$. Those can safely be neglected for the position, though.\\
\indent Finally, inverse Laplace transform of $\tilde{q}_{\rm III}$ can be neglected. The singularities are the same as for $\tilde{q}_{\rm II}$. For the singularities that do not depend on $\omega$, the associated residue is of order $1/\omega^2$ and no $\omega$ prefactor will appear when taking the time derivative. For the singularities that do depend on $\omega$, such a prefactor would appear, but as the residue is of the order $1/\omega^6$, it will not contribute either.\\
\indent We then consider that the time is the one elapsed in the linear region and the phase is the initial phase plus the one accumulated during the whole motion. This means doing the shifts $t \to t-t_0$ and $\varphi \to \varphi + \omega t_0$, with $t_0$ the time elapsed before the system entered the linear region. Combining all this in \eqref{app_eq_residueTheorem} and renaming the position in the linear region $q_{\rm lin}$ for clarity yield the solution in leading order in $1/\omega$: 
\begin{align}
    q_{\rm lin}(t) = &q_0 \cosh(\lambda_{\text{s}} (t-t_0)) \nonumber\\
    &+ \left( \frac{p_0}{\lambda_{\text{s}}} - \kappa  \sin({\varphi}+ \omega t_0) q_0\right) \sinh(\lambda_{\text{s}} (t-t_0)).
\end{align}
The momentum in the linear region is given by 
\begin{align}
    p_{\text{lin}}(t) &= \left(q_0 \lams +\kappa p_0 \sin(\omega t + \varphi) \right)\sinh(\lambda_{\text{s}} (t-t_0)) \nonumber\\
    &+\left( p_0 + \kappa \lams q_0 (\sin(\omega t+\varphi)-\sin({\varphi}+\omega t_0))\right) \nonumber\\
    &\ \ \ \cross\cosh(\lambda_{\text{s}} (t-t_0)).
\end{align}
The corresponding stability matrix is given by 
\begin{align}
    M(t) &= 
    \begin{pmatrix*}
        \frac{\partial q_{\text{lin}}(t)}{\partial q_0}     & \frac{\partial q_{\text{lin}}(t)}{\partial p_0}      \\
        \frac{\partial p_{\text{lin}}(t)}{\partial q_0}     & \frac{\partial p_{\text{lin}}(t)}{\partial p_0}        \\
    \end{pmatrix*}
\end{align}
with
\begin{align}
     \frac{\partial q_{\text{lin}}(t)}{\partial q_0} &=\cosh(\lambda_{\text{s}} (t-t_0))  \nonumber\\
     & \ \ \ \ \ - \kappa  \sin({\varphi}+ \omega t_0) \sinh(\lambda_{\text{s}} (t-t_0)),\\
     \frac{\partial q_{\text{lin}}(t)}{\partial p_0} &=\frac{1}{\lambda_{\text{s}}} \sinh(\lambda_{\text{s}} (t-t_0)), \\
     \frac{\partial p_{\text{lin}}(t)}{\partial q_0}  &= \lams \sinh(\lambda_{\text{s}} (t-t_0)) \nonumber\\
     &\mkern-60mu + \kappa  (\sin(\omega t + \varphi) -\sin({\varphi}+ \omega t_0)) \lams \cosh(\lambda_{\text{s}} (t-t_0)), \\
     \frac{\partial p_{\text{lin}}(t)}{\partial p_0} &= \cosh(\lambda_{\text{s}} (t-t_0))\nonumber\\
     & \ \ \ \ + \kappa \sin(\omega t + \varphi) \sinh(\lams (t-t_0)).
\end{align}

\section{Solution in the homoclinic region}
\label{app_homoclinic}
In this appendix we show how to approximately solve the evolution equation \eqref{eq_motion1} of the driven pendulum in the homoclinic region and calculate the associated stability matrix.

The starting point is Eq.~\eqref{eq_motion1}, rewritten here,
\begin{align}
    \ddot{q}(t) = \lambda_{\text{s}}^2 \sin q(t) + \kappa \omega \lambda_{\text{s}} \sin q(t) \cos(\omega t+\varphi) \,,
\end{align}
for the sake of convenience.
A first integral yields
\begin{eqnarray}
\frac{1}{2} \dot{q}^2(t) & = & \frac{p_1^2}{2} + \lambda_{\text{s}}^2 \cos q_1 - \lambda_{\text{s}}^2 \cos q(t) \nonumber \\ && + \kappa \omega \lambda_{\text{s}} \int_{t_1}^t \sin q(t') \cos( \omega t' + \varphi) \dot{q}(t') dt' \nonumber \\ 
& \simeq & \delta E_1 + 2 \lambda_{\text{s}}^2 \sin^2 [q(t)/2] \nonumber \\ && + \kappa \lambda_{\text{s}} \sin(\omega t + \varphi) \dot{q}(t) + O(\kappa/\omega)  \label{app_eq:qdot2}
\end{eqnarray}
with $q_1 = q(t_1)$, $p_1 = \dot{q}(t_1)$, and
\begin{eqnarray}
\delta E_1 & = & \frac{p_1^2}{2} - \lambda_{\text{s}}^2 ( 1 - \cos q_1) - \kappa \lambda_s \sin(\omega t_1 + \varphi) p_1 \sin q_1 \nonumber \\
& \simeq & \frac{1}{2}(p_1^2 - \lambda_{\text{s}}^2 q_1^2 ) - \kappa \lambda_{\text{s}} \sin(\omega t_1 + \varphi) p_1 q_1 + O(q_1^3)
\end{eqnarray}
the deviation, $\delta E_1 = E_1 - E_{\text{sep}}$, of the initial energy $E_1 = H(q_1,p_1,t_1)$ from the unperturbed separatrix energy $E_{\text{sep}} = \lambda_{\text{s}}^2$.
Using the generic scalings $\kappa \ll 1$ and $\kappa \omega \sim \lambda_{\text{s}}$ resulting from secular perturbation theory (see Appendix \ref{app_secular}), terms of the order $\sim \kappa / \omega$ effectively scale as $\sim \kappa^2 / \lambda_{\text{s}} \sim \lambda_{\text{s}}/\omega^2$ and are henceforth neglected.

Right on the separatrix and in absence of the driving, i.e., for $\delta E_1 = \kappa = 0$, Eq.~\eqref{eq_motion1} --- and hence also Eq.~\eqref{app_eq:qdot2} --- is equivalent to %\cThom{it seems like it's already in the approximation}
\begin{eqnarray}
\dot{q}(t) = \pm 2 \lambda_{\text{s}} \sin [q(t)/2] \,, \label{app_eq:qsepdot}
\end{eqnarray}
where the sign denotes the initial direction of motion near the unstable fixpoint at the origin, i.e., outwards from ($+$) or inwards to ($-$) the fixpoint.
This equation is exactly solved as
\begin{equation}
q(t) = 4 \arctan\left( \tan(q_1/4) e^{\pm\lambda_{\text{s}}(t-t_1)}\right)\,.
\end{equation}
Focusing on outward motion and neglecting terms scaling as $\sim q_1^3$, the unperturbed separatrix solution is thus written as
\begin{equation}
q_{\text{sep}}(t) = 4 \arctan \left[\text{sgn}(q_1)e^{\tau(t)}\right] \label{app_eq:qsep}
\end{equation}
where we define $\text{sgn}(q_1)= q_1/|q_1|$ and
\begin{equation}
\tau(t) = \lambda_{\text{s}} (t - t_1) - \log(4/|q_1|) \,. \label{app_eq:tau}
\end{equation}

Consider now a tiny deviation from the above separatrix solution. 
We thus also have a tiny value for the energy deviation from the separatrix and can safely assume $\delta E_1 \ll \lambda_{\text{s}}^2 \sin^2[q(t)/2]$ for all $t$, provided $q(t)$ starts sufficiently far away from the unstable fixed point and does not approach too closely the fixpoint on the other side of the separatrix arc.
Taking the positive square root of Eq.~\eqref{app_eq:qdot2} and performing a Taylor series expansion in $\delta E_1 /\{\lambda_{\text{s}}^2 \sin^2[q(t)/2]\}$ yields
\begin{eqnarray}
\dot{q}(t) & \simeq & 2 \lambda_{\text{s}} \sin[q(t)/2] + \kappa \lambda_{\text{s}} \sin(\omega t + \varphi) \sin q(t) \nonumber \\ && + \frac{\delta E_1}{2 \lambda_{\text{s}} \sin[q(t)/2]} - \frac{\delta E_1^2}{16 \lambda_{\text{s}}^3 \sin^3[q(t)/2]} \,, \label{app_eq:qdot}
\end{eqnarray}
where we neglect terms $\sim O(\delta E_1^3)$ that will turn out to scale quadractically either with $\kappa$ or with the initial phase-space deviation from the separatrix.
Making the ansatz
\begin{equation}
q(t) = q_{\text{sep}}(t) + \delta q(t)
\end{equation}
and using the linearizing expansions
\begin{eqnarray}
\sin[q(t)/2] & \simeq & \sin[q_{\text{sep}}(t)/2] + \cos[q_{\text{sep}}(t)/2] \frac{\delta q(t)}{2} \,, \\
\sin[q(t)] & \simeq & \sin[q_{\text{sep}}(t)] + \cos[q_{\text{sep}}(t)] \delta q(t)
\end{eqnarray}
together with the relations
\begin{eqnarray}
\sin[q_{\text{sep}}(t)/2] & = & \frac{\text{sgn}(q_1)}{\cosh \tau(t)} \,, \\
\cos[q_{\text{sep}}(t)/2] & = & - \tanh \tau(t)
\end{eqnarray}
that result from Eqs.~\eqref{app_eq:qsep} and \eqref{app_eq:tau}, %\cThom{is it a good idea to present it that way where it's a bit more involved?}
one obtains from Eq.~\eqref{app_eq:qdot}
\begin{eqnarray}
\delta \dot{q}(t) & \simeq & \left( - \lambda_{\text{s}} \tanh \tau(t) + \frac{\delta E_1}{8 \lambda_{\text{s}}} \sinh[2\tau(t)]\right) \delta q(t) \nonumber \\ && + \text{sgn}(q_1)\left(\frac{\delta E_1}{2 \lambda_{\text{s}}} \cosh \tau(t)  - \frac{\delta E_1^2}{16 \lambda_{\text{s}}^3} \cosh^3 \tau(t)\right) \nonumber \\ && - \kappa \lambda_{\text{s}} \sin(\omega t + \varphi) \left( 2 \,\text{sgn}(q_1) \frac{\sinh \tau(t)}{\cosh^2\tau(t)} \right. \nonumber \\ && \left. - \frac{\sinh^2 \tau(t) - 1}{\cosh^2 \tau(t)} \delta q(t) \right) \,, \label{app_eq:dqdot}
\end{eqnarray}
neglecting terms that scale as $O(\kappa^2)$, $O(\kappa/\omega)$, $O(\delta E_1^3)$, $O(\delta E_1^2 \delta q)$, and $O(\delta q^2)$.

Upon integration by parts, the term oscillating as $\kappa \sin(\omega t + \varphi)$ will give rise to another contribution scaling as $\sim \kappa / \omega$ that can be neglected for the determination of $\delta q(t)$.
The solution of the above equation for $\delta q(t_1) = 0$ is then given by
\begin{eqnarray}
\delta q(t) & = &\text{sgn}(q_1)\int_{t_1}^t dt' \frac{\exp\left( - \frac{\delta E_1}{8 \lambda_{\text{s}}^2} \cosh^2\tau(t')\right)}{\exp\left( - \frac{\delta E_1}{8 \lambda_{\text{s}}^2} \cosh^2\tau(t)\right)} \nonumber \\ && \times\left( \frac{\delta E_1 \cosh^2 \tau(t')}{2 \lambda_{\text{s}} \cosh\tau(t)} - \frac{\delta E_1^2 \cosh^4 \tau(t')}{16 \lambda_{\text{s}}^3\cosh\tau(t)}\right)
\end{eqnarray}
and can be further simplified by  neglecting terms scaling as $O(\delta E_1^4)$ upon expansion of the exponential functions.
This yields
\begin{eqnarray}
\frac{\delta q(t)}{\text{sgn}(q_1)} & = & \frac{\delta E_1 \lambda_{\text{s}}(t - t_1)}{4 \lambda_{\text{s}}^2\cosh\tau(t)} \left( 1 + \frac{\delta E_1}{16 \lambda_{\text{s}}^2}\left[2 \cosh^2\tau(t) - 3\right]\right) \nonumber \\ && + \frac{\delta E_1\left\{ \sinh[2\tau(t)] - \sinh[2\tau(t_1)]\right\}}{8 \lambda_{\text{s}}^2\cosh\tau(t)} \nonumber \\ && \times \left( 1 + \frac{\delta E_1}{8 \lambda_{\text{s}}^2} \left[ \cosh^2 \tau(t) - 2 \right] \right) \nonumber \\ && - \frac{\delta E_1^2\left\{\sinh[4\tau(t)] - \sinh[4\tau(t_1)] \right\}}{256 \lambda_{\text{s}}^4 \cosh \tau(t)} \,. ·\label{app_eq:dqt}
\end{eqnarray}

The propagation within the homoclinic region is considered to be accomplished as soon as the opposite end of the separatrix arc is reached.
We thus define the homoclinic propagation time $T$ such that $q_{\text{sep}}(t_1 + T) = 2\pi \,\text{sgn}(q_1) - q_1$.
With Eqs.~\eqref{app_eq:qsep} and \eqref{app_eq:tau} one has $\lambda T = 2\log(4/|q_1|)$ and hence $\tau(t_1 + T) = - \tau(t_1) = \log(4/|q_1|)$, from which one obtains 
\begin{equation}
\cosh \tau(t_1+T) \simeq \sinh \tau(t_1+T) \simeq 2 / |q_1| + O(q_1)
\end{equation}
in leading order in $q_1$.
Inserting this into Eq.~\eqref{app_eq:dqt} and keeping only terms in leading order in $q_1$ yields the rather simple expression 
\begin{equation}
\delta q(t_2) \simeq \frac{\delta E_1}{\lambda_{\text{s}}^2 q_1}( 1 + O(q_1^2))
\end{equation}
for the position deviation from the separatrix solution at final time $t_2 = t_1 + T$.
The corresponding final momentum deviation, given by Eq.~\eqref{app_eq:dqdot} evaluated at $t_2 = t_1 + T$, is
\begin{equation}
\delta \dot{q}(t_2) \simeq - \frac{\delta E_1}{\lambda_{\text{s}} q_1} - \kappa \lambda_s q_1 \sin(\omega t_2 + \varphi)
\end{equation}
in leading order, neglecting terms that scale as $\delta E_1^2$.
The above two expressions further simplify if we neglect the deviation of the initial phase-space point from the separatrix and hence consider $q_1 = \pm q_c$ and $p_1 = \pm 2 \lambda_{\text{s}} \sin(q_c/2) \simeq \pm \lambda_{\text{s}} q_c$, with $q_c$ the crossover scale separating the linear from the homoclinic region.
We then have $\delta E_1 \simeq - \kappa \lambda_{\text{s}}^2 \sin(\omega t_1 + \varphi) q_c^2$
and thus
\begin{eqnarray}
\delta q(t_2) & \simeq & \mp \kappa q_c\sin(\omega t_1 + \varphi) \,, \label{app_eq:dq2}\\
\delta \dot{q}(t_2) & \simeq & \mp \kappa \lambda_{\text{s}} q_c\sin(\omega t_2 + \varphi)\,. \label{app_eq:dp2}
\end{eqnarray}

The associated stability matrix can be calculated by evaluating the effect of a tiny initial deviation with respect to this particular reference trajectory, at fixed total propagation time $T = 2 \log(4/q_c) / \lambda_{\text{s}}$.
Setting $q_1 = q_c + \delta q_1$ and $p_1 = \lambda_{\text{s}} q_c + \delta p_1$ (we henceforth restrict the discussion to the case $q_1 > 0$ without loss of generality), the separatrix solution \eqref{app_eq:qsep} is advanced by
\begin{equation}
\delta q_{\text{sep}}(t_2) = q_{\text{sep}}(t_2) - (2\pi - q_c) \simeq \delta q_1
\end{equation}
at the end of the propagation.
With Eq.~\eqref{app_eq:qsepdot}, the corresponding advancement of the time derivative of the separatrix solution is obtained as
\begin{eqnarray}
\delta \dot{q}_{\text{sep}}(t_2) & = & 2 \lambda_{\text{s}} \sin[q_{\text{sep}}(t_2)/2] - 2 \lambda_{\text{s}} \sin(q_c/2) \nonumber \\ &\simeq& -\lambda_{\text{s}} \delta q_1\,.
\end{eqnarray}

Additional shifts of the final position and momentum variables result from Eqs.~\eqref{app_eq:dqdot} and \eqref{app_eq:dqt}.
Using 
\begin{eqnarray}
\cosh \tau(t_1) \simeq - \sinh\tau(t_1) & \simeq & \frac{2}{q_c}\left( 1 - \frac{\delta q_1}{q_c}\right) \,, \\
\cosh \tau(t_2) \simeq \sinh\tau(t_2) & \simeq & \frac{2}{q_c}\left( 1 + \frac{\delta q_1}{q_c}\right) \,,
\end{eqnarray}
and
\begin{eqnarray}
\delta E_1 & \simeq & \lambda_{\text{s}} q_c (\delta p_1 - \lambda_{\text{s}} \delta q_1) - \kappa \lambda_{\text{s}} q_c\sin(\omega t_1 + \varphi) \nonumber \\ && \times [\lambda_{\text{s}} (q_c + \delta q_1) + \delta p_1] \,,
\end{eqnarray}
all expanded in linear order in $\delta q_1, \delta p_1$, one obtains from Eq.~\eqref{app_eq:dqt}
\begin{equation}
\delta q(t_2) \simeq \frac{\delta p_1}{\lambda_{\text{s}}} - \delta q_1 - \kappa \sin(\omega t_1 + \varphi)\left( q_c + \frac{\delta p_1}{\lambda_{\text{s}}}\right)
\end{equation}
in leading order in $q_c$ (more precisely, up to terms that scale as $\sim q_c^2 \log q_c$).
Inserting this expression into Eq.~\eqref{app_eq:dqdot} yields the corresponding final momentum shift
\begin{eqnarray}
\delta \dot{q}(t_2) & \simeq & - \kappa \sin(\omega t_1 + \varphi) ( \delta p_1 + \lambda_{\text{s}} \delta q_1) \nonumber \\ && - \kappa \sin(\omega t_2 + \varphi)( \lambda_{\text{s}} q_c - \delta p_1) \,,
\end{eqnarray}
again in leading order in $q_c$.

The final position and momentum deviations from the reference expressions \eqref{app_eq:dq2} and \eqref{app_eq:dp2} are then obtained as
\begin{eqnarray}
\delta q_2 & = & \delta q_{\text{sep}}(t_2) + \delta q(t_2) - \left.\delta q(t_2)\right|_{\delta q_1 = \delta p_1 = 0} \nonumber \\
& \simeq & [1 - \kappa \sin(\omega t_1 + \varphi)]\frac{\delta p_1}{\lambda_{\text{s}}} \,, \\
\delta p_2 & = & \delta \dot{q}_{\text{sep}}(t_2) + \delta \dot{q}(t_2) - \left.\delta \dot{q}(t_2)\right|_{\delta q_1 = \delta p_1 = 0} \nonumber \\
& \simeq & - [ 1 + \kappa \sin(\omega t_1 + \varphi)] \lambda_{\text{s}} \delta q_1 \nonumber \\ && + \kappa [\sin(\omega t_2 + \varphi) - \sin(\omega t_1 + \varphi)] \delta p_1
\end{eqnarray}
in leading order in $q_c$.
This yields the stability matrix
\begin{equation}
M = \left( \begin{array}{cc} \frac{\delta q_2}{\delta q_1} & \frac{\delta q_2}{\delta p_1} \\ \frac{\delta p_2}{\delta q_1} & \frac{\delta p_2}{\delta p_1}\end{array} \right) \simeq \left( \begin{array}{cc} 0 & (1 - \kappa_1)/\lambda_{\text{s}} \\ - (1 + \kappa_1)\lambda_\text{s} & \kappa_2 - \kappa_1 \end{array} \right)
\end{equation}
where we use the abbreviations $\kappa_j = \kappa \sin(\omega t_j + \varphi)$ for $j=1,2$.

\section{Crossover between the linear and the homoclinic region} \label{app_sec_melnikov}

In this appendix, we use the Melnikov-Arnold integral \cite{Arnold_1964, Melnikov_1962, Chirikov_1979, Lichtenberg_Lieberman_2013} to derive a lower bound for the crossover position scale $q_c$ that formally separates the linear from the homoclinic region in our analytical treatment of the driven pendulum problem.
We calculate, for this purpose, the energy gain $\Delta H$ of a trajectory that closely follows the broken separatrix structure in the presence of a weak perturbation from integrability.
This energy gain is given by 
\begin{eqnarray}
\Delta H & = & \int_{t_1}^{t_2} \frac{d}{d t} H[q(t),p(t),t] dt \nonumber \\
& = & - \kappa \omega^2 \lambda_{\text{s}} \int_{t_1}^{t_2} \cos[q(t)] \sin(\omega t + \varphi) dt \,,
\label{eq:Egain}
\end{eqnarray}
using
$\frac{d}{d t} H[q(t),p(t),t] = \frac{\partial}{\partial t} H[q(t),p(t),t]$ with $H$ given by Eq.~\eqref{eq_hamiltonian}.

In leading order in the perturbation from integrability, 
$q(t)$ in the above expression \eqref{eq:Egain} can be approximately substituted by the unperturbed separatrix solution $q_{\text{sep}}(t)$ given by Eq.~\eqref{app_eq:qsep} (where we use, without loss of generality, $\textrm{sgn}(q_1) = 1$).
The latter is rewritten as
\begin{equation}
q_{\text{sep}}(t) = 4 \arctan e^{\lambda_\text{s} (t - t_\pi)}
\end{equation}
where $t_\pi$ is the evolution time at which $q_{\text{sep}}(t_\pi) = \pi$.
Considering the limits $t_1-t_\pi \to -\infty$ and $t_2 - t_\pi \to \infty$, Eq.~\eqref{eq:Egain} is then approximately evaluated as
\begin{eqnarray}
\Delta H & \simeq & - \kappa \omega^2 \lambda_{\text{s}} \int_{-\infty}^{\infty} \cos(4 \arctan e^{\lambda_\text{s} t'}) \nonumber \\ && \times \sin[\omega (t_\pi + t') + \varphi] dt'\nonumber \\
& = & - \frac{\kappa \omega^2}{4 i} \left\{ \left[ A(\omega/\lambda_{\text{s}}) + A^*(-\omega/\lambda_{\text{s}})\right] e^{i(\omega t_\pi + \varphi)} \right. \nonumber \\ && \left. - \left[ A^*(\omega/\lambda_{\text{s}}) + A(-{\omega}/\lambda_{\text{s}})\right]e^{-i(\omega t_\pi + \varphi)} \right\} \,,
\end{eqnarray}
where we introduce by 
\begin{equation}
A(\Omega) = \int_{-\infty}^\infty \exp[i(4 \arctan e^t + \Omega t)] dt
\end{equation}
the Melnikov-Arnold integral \cite{Melnikov_1962,Arnold_1964,Chirikov_1979}.
Using 
\begin{equation}
\exp(4 i \arctan e^t) = \left(\frac{1+ i e^{t}}{1 - i e^{t}}\right)^2
\end{equation}
and applying residue calculus, one obtains
\begin{equation}
A(\Omega) = -\frac{4 \pi \Omega}{\sinh \pi \Omega} e^{-\pi \Omega / 2} \,.
\end{equation}
This yields the expression
\begin{eqnarray}
\Delta H &\simeq& \frac{2\pi\kappa\omega^3 \sin(\omega t_\pi + \varphi)}{\lambda_{\text{s}}\sinh(\pi \omega/2 \lambda_\text{s})} 
\end{eqnarray}
for the asymptotic energy gain along the broken separatrix structure.

Using $\omega \gg \lambda_{\text{s}}$ and performing an average over the specific driving phase $\omega t_\pi + \varphi$ arising at $t = t_\pi$, the mean scale for the energy deviation along the separatrix is yielded as
\begin{equation}
\overline{\Delta H} = \left(\overline{\Delta H^2}\right)^{1/2} \simeq \frac{\sqrt{2}\pi \kappa \omega^3}{\lambda_{\text{s}}\sinh(\pi \omega/2 \lambda_\text{s})} \,. \label{eq:Egain_av}
\end{equation}
This energy gain or loss can be approximately translated into a mean momentum deviation from the unperturbed separatrix trajectory, $\delta p(t_2) = \dot{q}(t_2) - \dot{q}_{\text{sep}}(t_2)$, evaluated for $q(t_2) = 2\pi- q_c$ at the end of the homoclinic region.
Using $\Delta H \simeq [\dot{q}^2(t_2) - \dot{q}^2_{\text{sep}}(t_2)]/2$ with $\dot{q}_{\text{sep}} \simeq \lambda_{\text{s}} q_c$, one obtains $\Delta H \simeq \lambda_\text{s} q_c \delta p(t_2)$ in leading order in $\delta p$.
The requirement $|\delta p(t)| \ll \lambda_{\text{s}} q_c$, which translates closeness to the separatrix structure in the homoclinic region, then yields the inequality $|\Delta H| \ll (\lambda_\text{s} q_c)^2$,
from which follows, with Eq.~\eqref{eq:Egain_av},
\begin{equation}
q_c^2 \gg \frac{\overline{\Delta H}}{\lambda^2} = \frac{\sqrt{2}\pi \kappa (\omega/\lambda_\text{s})^3}{\sinh(\pi \omega/2 \lambda_\text{s})} \,.
\end{equation}
This latter inequality constitutes a lower bound for the crossover position scale $q_c$.

%\section*{References}
\bibliography{bibliography}

\end{document}